\documentclass{JHEP3}
\usepackage{latexsym,bm,amsmath,amssymb,amsfonts}
\usepackage{epsfig,graphics,graphicx,mathrsfs}
\usepackage{slashed}
\usepackage{cite}
\usepackage[latin1]{inputenc}
\long\def\comment#1{ }

  \newcommand{\abar}{\bar{\alpha}_s}

  \newcommand{\mcal}{\mathcal}
  
  \newcommand{\rmd}{{\rm d}}

\newcommand{\mcA}{{\mathcal{A}}}
 \newcommand{\beq}{\begin{eqnarray}}
  \newcommand{\eeq}{\end{eqnarray}}

\newcommand{\qint}{{\int \frac{d^2q}{\pi q^2}}}
 \newcommand{\melint}{{\int \frac{d \gamma}{2\pi \rm i}}}

\newcommand{\kprime}{{|\pmb{k}+\pmb{q}|}}

\title{Non-linear evolution in CCFM: The interplay between coherence 
and saturation}

\author{Emil Avsar$^a$ and Anna M. Stasto$^{a,b,c}$ \\
\!\!$^{a}$104 Davey Lab, Penn State University, University Park, 16802 PA,  USA\\  
\!\!$^b$RIKEN Center, Brookhaven National Laboratory, Upton, NY 11973, USA\\
\!\!$^c$Institute of Nuclear Physics Polish Academy of Sciences, Cracow, Poland\\
E-mail:  \email{eavsar@phys.psu.edu,astasto@phys.psu.edu}}

\abstract{
We solve the CCFM equation numerically in the presence of a boundary condition 
 which effectively  incorporates the  non-linear dynamics. We retain the full dependence
 of the unintegrated gluon distribution on the coherence scale, and  extract the saturation 
momentum. The resulting saturation scale  is  a function of both rapidity and  the coherence momentum. In Deep Inelastic Scattering this will lead to a dependence of the saturation scale on the photon virtuality in addition to the usual  $x_{Bj}$ dependence.
 At asymptotic energies the interplay between the 
perturbative non-linear physics, and that of the QCD coherence, leads to an interesting 
and novel dynamics where the saturation momentum itself eventually saturates.
We also investigate various implementations of the ''non-Sudakov'' form factor. It is  shown that  the non-linear dynamics leads to almost identical results for different form factors.
Finally, different choices of the scale of the running coupling are analyzed and implications for the phenomenology are discussed.
}

\begin{document}
\section{Introduction}

The linear QCD evolution equations at small-$x$, such as BFKL \cite{BFKL} or CCFM \cite{CCFM1,CCFM2,CCFM3,Marchesini:1995ia,Marchesini:1991sg}, lead 
to a very rapid growth of the gluon density. Once the density is too large however, 
one needs to take into account the  non-linear effects. The BFKL evolution is modified in the 
high density limit by the so-called Color Glass Condensate (CGC) \cite{McLerran:1993ni,McLerran:1993ka} (for a review see \cite{CGCreviews}) formalism
where the strong color fields are treated semi--classically, and where their
presence leads to the saturation of the gluon emission rate\footnote{This 
implies that the gluon density keeps growing also in the saturated regime
but the growth is logarithmic in $x$ as opposed to power-like.}. The BFKL
evolution equation is then generalized to either the Balitsky-Kovchegov (BK) \cite{Balitsky:1995ub, Kovchegov:1999yj}
equation or to the more complete JIMWLK \cite{JKLW, CGC} equation. The energy dependent 
momentum scale characterizing the transition from the weak to strong field
regimes is called the saturation momentum and is denoted $Q_s(x)$. 

What makes CCFM different from the standard BFKL-type picture of the small-$x$ regime is 
the property of the so-called coherence of emissions. This is
quantum mechanical in origin and comes from the interference in the 
emission of gluons from different partons. This leads to what 
is termed angular ordering which, as a fundamental feature of QCD, 
is present in both time-like and space-like evolution as well as at 
both small and large $x$. In CCFM
the effects of angular ordering are encoded in an additional momentum scale, usually
denoted $p$, so that the unintegrated gluon density $\mcA$ now depends on 
three variables: Bjorken $x$, the momentum $k$, and the additional 
momentum $p$. We will denote the unintegrated gluon density by  $\mcA(x,k,p)$. In comparison, the BFKL gluon 
density, and similarly the density in the CGC formalism, depends only on $x$ and $k$, that is
$\mcA=\mcA(x,k)$. 

The saturation scale $Q_s$ is the momentum at which the gluons in the 
hadron start to overlap  with each other.  This leads to the increased rate for the recombination of gluons.
$Q_s$ can be defined as the momentum for which
$\mcA$ is a given constant, say of the order of 1. This is a natural definition
since the gluon density can be defined as a phase space occupation number\footnote{
The exact relation between $\mcA$ and the occupation number depends 
on the precise definition of the former.}
of gluons per unit rapidity with given transverse momentum $k$. This 
automatically implies that $Q_s$ is $x$ dependent. However,  
it is then clear that in CCFM the same definition immediately leads to 
an extra scale dependence of $Q_s$, since now the requirement of constant occupation 
number, $\mcA(x,Q_s,p)= {\rm const}$, leads to $Q_s \equiv Q_s(x,p)$. One of our main objectives
in this paper is to study this novel scale dependence of $Q_s$ by 
numerically solving the CCFM equation in the presence of a boundary 
condition enforcing non-linear corrections. 

The additional momentum variable $p$ in CCFM is determined by 
the kinematics of the hard scattering. More precisely, the hard scattering 
determines the maximal angle of the allowed emissions in accordance 
with angular ordering. Conventionally one uses the squared angle $\xi = q_\perp^2/q_l^2$
as the angular variable, and for example in a DIS process with $n$ gluon emissions, 
angular ordering implies that $\xi_1 < \xi_2 < \cdots < \xi_n < \bar{\xi}$
where $\bar{\xi}$ is the maximal angle determined by the hard scattering. 
The transverse momentum $p$ is then defined 
via $\bar{\xi}=p^2/(x_n^2s)$ where $x_n$ is the energy fraction of 
the gluon entering the hard scattering, and $s$ is the cms energy. 
The variable $p$ is related to $Q$ in DIS, roughly
$\bar{\xi} \simeq Q^2/(x^2s)$. 
This means that the saturation scale from CCFM in DIS kinematics will depend on the 
photon virtuality $Q$ in addition to  the Bjorken $x$.
 In the usual picture of the small-$x$ physics, however, $Q_s$ is an intrinsic property 
of the target hadron, and is independent of the probe and its virtuality.
The difference now is that the gluon occupation number in the hadron 
depends additionally on the probe.  
That $Q_s$ is not a property of the target alone but also depends 
on the probe and its virtuality is certainly what one would generally expect from
quantum mechanics. In the end $Q_s$ is determined by the scattering process, 
to which both the probe and the target enters, and therefore we should expect 
$Q_s$ to depend on both. In that sense the extra scale dependence of $Q_s$ 
is most natural. 

Unlike BFKL, the proper non-linear generalization of CCFM 
is not know yet, and we shall not present it here either. 
In \cite{Avsar:2009pv, Avsar:2009pf}, however, a method 
was proposed which effectively implements saturation and unitarity 
corrections within a generic linear evolution equation, and the method 
was in particular applied to BFKL and CCFM. The idea is simply based 
on enforcing a boundary condition in the linear evolution which 
prevents the gluon phase--space occupation numbers to grow too large, 
and the procedure is an extension of a strategy originally introduced in connection
with analytic studies of BFKL evolution in the presence of saturation 
effects \cite{Mueller:2002zm, Iancu:2002tr}. The method was demonstrated 
to work very well for BFKL in \cite{Avsar:2009pv} where the solutions 
of BFKL in the presence of the saturation boundary condition
were compared to the solutions of BK, and
it was shown that the method reproduces successfully the results from BK for any $x$ 
and for both fixed and running $\abar$. 

The method was then applied to CCFM in \cite{Avsar:2009pf} and $Q_s$ was 
extracted. However, in order to simplify the numerical procedure, only 
the case $k \ll p$ was studied in which case one can neglect 
the dependence on $p$ so that $Q_s$ depends only on $x$. This is the 
region that leads to another formulation of CCFM called the Linked Dipole
Chain (LDC) model \cite{Andersson:1995ju} which has some advantages compared to CCFM such 
as projectile-target symmetry. 
In this paper, we will keep the full $p$ 
dependence of the evolution, and we will show that this dependence,
which as mentioned above is related to coherence in the emissions, 
combined with the 
physics of saturation gives rise to a very interesting and novel dynamics which 
eventually leads to the saturation of the saturation momentum $Q_s$ itself.
This happens when the solution to $\mcA$ ''stalls'' in some region of $k$, 
meaning that the evolution in $Y$ is almost completely stopped.  
This behavior is in contrast to the standard picture of the non-linear dynamics
where $Q_s$ grows indefinitely with $Y$. 

Generally we shall see that the shape of $Q_s$ and that of $\mcA$ 
close to the saturation regime is in CCFM rather different than in BFKL.
Also some characteristic features of the standard small-$x$ evolution 
such as geometric scaling \cite{Stasto:2000er} is modified 
in CCFM. On the practical level, however, it seems that the new behavior
of $Q_s$, whereby it eventually saturates as a function of $Y$ for given $p$,
sets in very late, especially with a running coupling, so that it seems it would be 
be very hard to observe at collider experiments.  On the other hand it will 
still be important to take into account the variation of $Q_s$ with $p$ 
also for phenomenological values of $Y$, especially when $p$ is small. 

In this study, we restrict ourselves to the version of CCFM containing only 
the ''hard'' emissions associated with the $1/z$ pole in the gluon splitting function. 
In addition to these, in CCFM there are also  the ''soft'' emissions which 
are associated with the opposite $1/(1-z)$ pole\footnote{In addition to the 
singular pieces of the splitting function, the non-singular pieces are 
usually included in phenomenological applications, such as in the CASCADE 
Monte Carlo \cite{Jung:2000hk}. While these terms may obviously be important in practice it should
be noted that their inclusion is somewhat ad hoc since they were not properly included
in the original formulation of CCFM.}. Within the accuracy of the formulation
of CCFM however, these emissions are exactly ''probability conserving'', meaning 
that they are exactly compensated by the associated virtual corrections encoded 
in the so-called ''Sudakov'' form factor. Thus at least formally, it should make 
no difference for the solution of $\mcA$ whether one includes the soft emissions
or not. The difference would be only in the exclusive final state, a correct 
description of which requires the soft emissions. 
In practice, however, such a formal statement carries little significance, 
and the inclusion of soft emissions can have substantial effects on $\mcA$. Thus 
it is our aim to also include the soft emissions in our procedure. 
This is, however, somewhat challenging numerically due to the fact that 
the real and virtual terms are separately divergent, and so one must introduce 
a cut-off which complicates the numerical procedure. Therefore the inclusion 
of the potentially important soft emissions and the associated Sudakov 
form factor is postponed to a future work. 

The $1/z$ emissions are accompanied by virtual corrections which are encoded 
in the so-called ''non-Sudakov'' form factor. One can in the literature find 
different expressions for this form factor.   In this 
paper we shall study both of the common expressions for the form factor.
We shall see that, generally, the 
different choices of the non-Sudakov form factors lead to rather different 
results in the linear evolution, which is due to the reason that the differences 
which occur in the softer region are enhanced unless there is some mechanism 
to cut off the growth there. As we will demonstrate, once non-linear evolution is included via the 
saturation boundary condition, then  the differences are to a large degree removed. The small differences remaining when $Y$ is extended to very large values can 
can be removed by scale factors independent of $Y$ and $p$. Thus saturation introduces a certain universality in the evolution as was already observed in  \cite{Avsar:2009pf}. 

Related to this, it is important to check the sensitivity of the
results to different prescriptions of the boundary condition. Generally, 
one can consider different boundary conditions or, for a given type
of boundary, different values of the boundary parameters. In \cite{Avsar:2009pf} 
it was reported that the different choices lead to different normalizations
of the solutions, without altering the shape or the $Y$ dependence. Since 
we here keep the $p$ dependence, the question is more subtle. 
We will show that the differences are small and that they can generally be removed by 
$Y$ and $p$ independent scale factors. Some differences can remain 
in the region where $Q_s$ starts to saturate (the region where $\mcA$ ''stalls''), 
but what we find is that this behavior itself, \emph{i.e} 
the saturation of  $Q_s$,  is 
independent of the precise application of the boundary. What changes
is simply at what $Y$, for a given $p$, this behavior sets in. 
To know the exact details of the behavior in the saturated 
regime one would need to fully formulate the proper non-linear generalization
of CCFM which is beyond the scope of this work.  
It seems, however, rather reasonable to expect that this interesting 
dynamics where, as a result of the combination of saturation and coherence, 
$Q_s$ eventually saturates will remain true also once the proper non-linear 
generalization of CCFM is known. 

The structure of the paper is the following: in the next section we present a brief overview of the CCFM formalism and show the numerical results for the solution in the case when the full
dependence on the scale $p$ is included. We also discuss different versions of the running coupling prescriptions. In section~3  we provide semi-analytical results for the case of CCFM with the boundary condition. In particular, we discuss the dependence of the saturation scale on the rapidity and $p$. In section~4 we briefly describe the method of implementation of the saturation boundary. In section~5 we present the complete numerical analysis of CCFM in the presence of the saturation boundary. We analyze different prescriptions for the non-Sudakov form factors and the running  coupling, and  we extract the rapidity and $p$ dependence of the saturation scale. Finally, in section~6 we state our conclusions and discuss the relevance of the results for the phenomenology.

\section{Overview of CCFM}

Our aim in this section is to provide a brief overview of CCFM, focusing 
on the points which shall be relevant for our analysis. 
It is possible to find different versions of the non-Sudakov
form factor in the literature  but the origin of this form factor is hardly ever discussed. 
It will therefore be important to recall the derivation of this
form factor and understand the motivation behind the formulas. Moreover,
we shall in addition to the standard choice of the scale of the 
running coupling also consider a different choice.
The most comprehensive
overview is to be found in the original papers in \cite{CCFM1,CCFM2,CCFM3,Marchesini:1995ia, Marchesini:1991sg}, and a 
simplified but rather detailed discussion on CCFM is also offered in \cite{Avsar:2009pf}. 

\begin{figure}[t]
\begin{center}
\includegraphics[angle=0,width=0.45\textwidth]{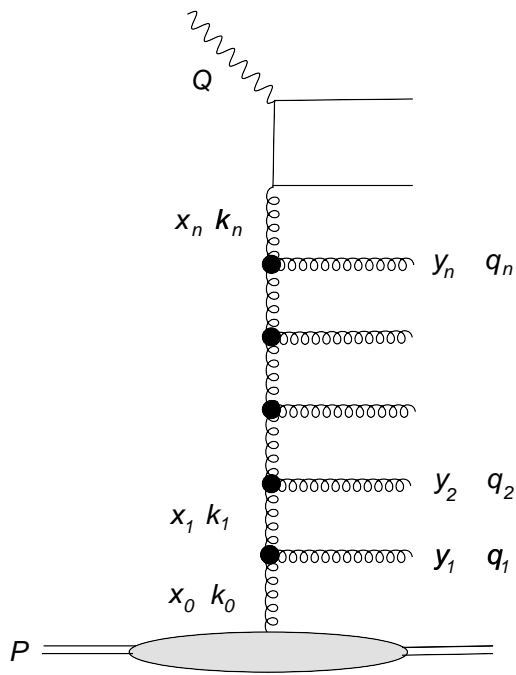}
\end{center}
\caption{\label{fig:ladder} The kinematics of the gluon ladder. The energy fraction of the 
real $s$-channel and virtual $t$-channel gluons are denoted by $y_i$ and $x_i$ respectively while 
their respective transverse momenta are denoted by $q_i$ and $k_i$.
}
\end{figure}

The gluon ladder generated by the CCFM evolution is represented in figure~\ref{fig:ladder}. 
In CCFM the successive emissions along the ladder are ordered in their angle. 
As mentioned in the introduction, it is customary to define $\xi = q_\perp^2/q_l^2$ as the angular variable (with 
the direction of the incoming proton defining the longitudinal direction), and 
angular ordering implies that $\xi_i > \xi_{i-1}$. Additionally there is 
a maximum angle $\bar{\xi}$, set by the kinematics of the hard scattering, 
which limits all emissions.  
  
In CCFM, the emissions are divided into hard and soft emissions, and the importance
of this classification is that it offers significant simplifications in the 
formulation. The precise definition is that the hard emissions 
are those emissions ordered in \emph{both} energy and angle, while for each soft
emission there exists another emission with bigger angle and larger energy fraction.
It follows that, the hard emissions are those associated with the $1/z$ pole 
in the gluon splitting function while the soft emissions are associated 
with the opposite $1/(1-z)$ pole. Following the above definition one 
can associate with each hard emission $k$ a cluster of soft emissions $S_k$
which satisfy $\xi_{k-1} < \xi < \xi_k$ and 
$y < y_k$ where $\xi_{k-1}$ is the angle of the previous hard emission. 
The label ''soft'' has a two-fold meaning. Firstly the fact that each 
of these emissions takes away very  
little energy, because of the $1/(1-z)$ pole, and secondly that
all soft emissions in the set $S_k$ satisfy $q < q_k$.  
The classification of the emissions as being hard or soft is also related 
to another classification of the ladder in terms of so-called  $k_\perp$--changing and 
$k_\perp$--conserving emissions respectively. What is meant by this is that there are certain 
real emissions which leave the virtual propagator momenta unchanged in an approximate sense. 
This latter classification is crucial for deriving the 'standard' version of 
CCFM which is used in all practical applications.

\subsection{Integral equations and the virtual form factors}

The CCFM integral equation, including both the soft and the hard emissions,
is given by 
\begin{multline}
\mathcal{A}(x, k, p) \; = \; \abar \int_x^1 dz
\int \frac{\rmd ^2\bar{q}}{\pi \bar{q}^2} \,
\theta (p - z\bar{q}) \, \Delta_s(p,z\bar{q})
\left ( \frac{\Delta_{ns}(z,k, q)}{z} + \frac{1}{1-z} \right )
\mathcal{A}\left(\frac{x}{z}, k', \bar{q}\right) \; .
\label{eq:ccfminteq1}
\end{multline}
The momentum variable $p$ is defined via $\bar{\xi} = p^2/(x_n^2s)$, and $k' = |\pmb{k} + (1-z)\bar{\pmb{q}}|$.
The momentum $\bar{q}$ is the rescaled momentum of the real gluon, and is related 
to the true momentum $q$ by $\bar{q} = q/(1-z)$. In line with all the approximations 
made in order to derive this equation one can for the $1/z$ pole set $\bar{q}=q$ 
(which is what we have already done in $\Delta_{ns}(z,k,q)$).
Here $\Delta_s$ is the Sudakov
form factor which screens the singularity of the $1/(1-z)$ pole, while 
$\Delta_{ns}$ is the ''non-Sudakov'' form factor which depends also on the 
 momenta $k$ and it corresponds in BFKL to the gluon Regge factor.
   The rescaled coupling constant $\abar$ is defined as $\alpha_s N_c/\pi$.
   
\begin{figure}[t]
\begin{center}
\includegraphics[angle=0, width=0.6\textwidth]{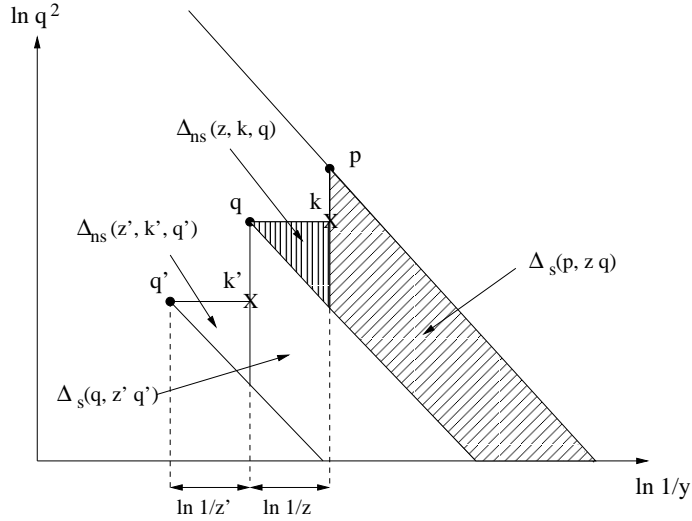}
\caption{Graphical representation of equation \protect \eqref{eq:ccfminteq1}. 
The horizontal axis denotes the inverse energy fraction while the vertical axis 
denotes the transverse momentum, all in logarithmic scale. The shaded regions 
show the regions over which the form factors receive contributions. 
The real emission \protect $q$ is represented by a dot, while the virtual gluons \protect $k$ and \protect $k'$
are represented by crosses. The position of \protect $p$ is also indicated. The diagonal 
lines are lines of constant $\xi$. }
\label{fig:phasespace}
\end{center}
\end{figure}
 
In figure~\ref{fig:phasespace} we represent pictorially the components of equation \eqref{eq:ccfminteq1}.
Here the horizontal axis represents the inverse energy fraction of the gluons, the further to 
the right a gluon is located the less energy it has, while the vertical axis represents the transverse 
momenta, the higher up a emission is located the larger its transverse momentum. The maximal angle 
$\bar{\xi}$ is represented by drawing a line of constant angle from the point $(x_n,p)$; 
this is the diagonal line going through the black dot representing $p$ ($p$ itself is not the 
momentum of any gluon (real or virtual) but we still denote its location by a dot in the figure). 
In this particular case, $k < p$. The ''Sudakov'' form factor $\Delta_s(p,z\bar{q})$ is then given by 
exp$(-\abar S)$ where $S$ is the phase space area which is the diagonally shaded region in 
the figure. This is the region where further emissions would have been possible. 
Similarly the ''non-Sudakov'' form factor $\Delta_{ns}$ is given by exp$(-\abar A)$ where $A$ 
is the area of the vertically shaded region.  In the figure we also show 
the next backward step in the evolution which is obtained from \eqref{eq:ccfminteq1} when 
$\mcA$ in the RHS is expanded iteratively. As we can see from the figure, 
the region $S$ contributing to $\Delta_{s}$ diverges when extended down indefinitely. Thus this divergence is cut by a 
horizontal line in the figure, which means a soft cut-off in the transverse momentum. Notice that 
this divergence is the one arising from the $1/(1-z)$ pole because as $z \to 1$, the gluon 
carries infinitely small energy fraction, and in the figure it would then be located 
precisely in the region where the diagonally shaded region is extended down indefinitely 
(the fact that this region is bounded by the diagonal lines is due to angular ordering). 

The soft emissions are such that they are exactly compensated by the factors $\Delta_s$. 
This means that the regions over which the Sudakov form factors receive contribution are
precisely the regions to which the soft emissions are confined, \emph{i.e.} 
the clusters $S_k$ mentioned above. Summation 
over all possible soft emissions in each $S_k$ leads to a factor
exp$(+\abar S_k)$. Each such factor then multiplies to unity with the corresponding Sudakov form factor exp$(-\abar S_k)$. 
One is consequently left with the hard emissions only, and the integral equation reads 
\begin{eqnarray}
\mathcal{A}(x, k, p) = \abar \int_x^1 \frac{\rmd z}{z}
\int \frac{\rmd ^2 q}{\pi q^2}\, \theta (p - z q) \, \Delta_{ns}(k,z,q) \,
\mathcal{A}\left(\frac{x}{z}, k', q\right) \; .
\label{eq:ccfminteq2}
\end{eqnarray}
This equation is thus formally equivalent to  \eqref{eq:ccfminteq1}. 
In practice, however, one can expect to find different solutions. The numerical implementation
of the soft emissions faces the difficulty that one must cut the $z\to 1$ singularity by 
a momentum cut-off as in figure~\ref{fig:phasespace}. This introduces a numerical uncertainty 
which complicates the procedure. Therefore, even though their inclusion is important and 
interesting, we will postpone their treatment to a future work
and in the present concentrate only on equation \eqref{eq:ccfminteq2}. 

\begin{figure}[t]
\begin{center}
\includegraphics[angle=0, width=0.6\textwidth]{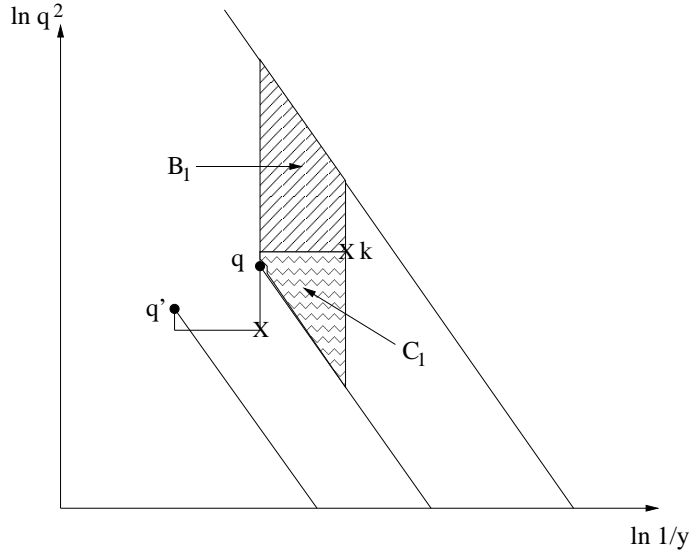}
\end{center}
\caption{\label{fig:nonSud} Illustration of the regions contributing to the definition 
of the non-Sudakov form factor $\Delta_{ns}$. In this case $k \geq q$.
}
\end{figure}

\begin{figure}[t]
\begin{center}
\includegraphics[angle=0, scale=0.45]{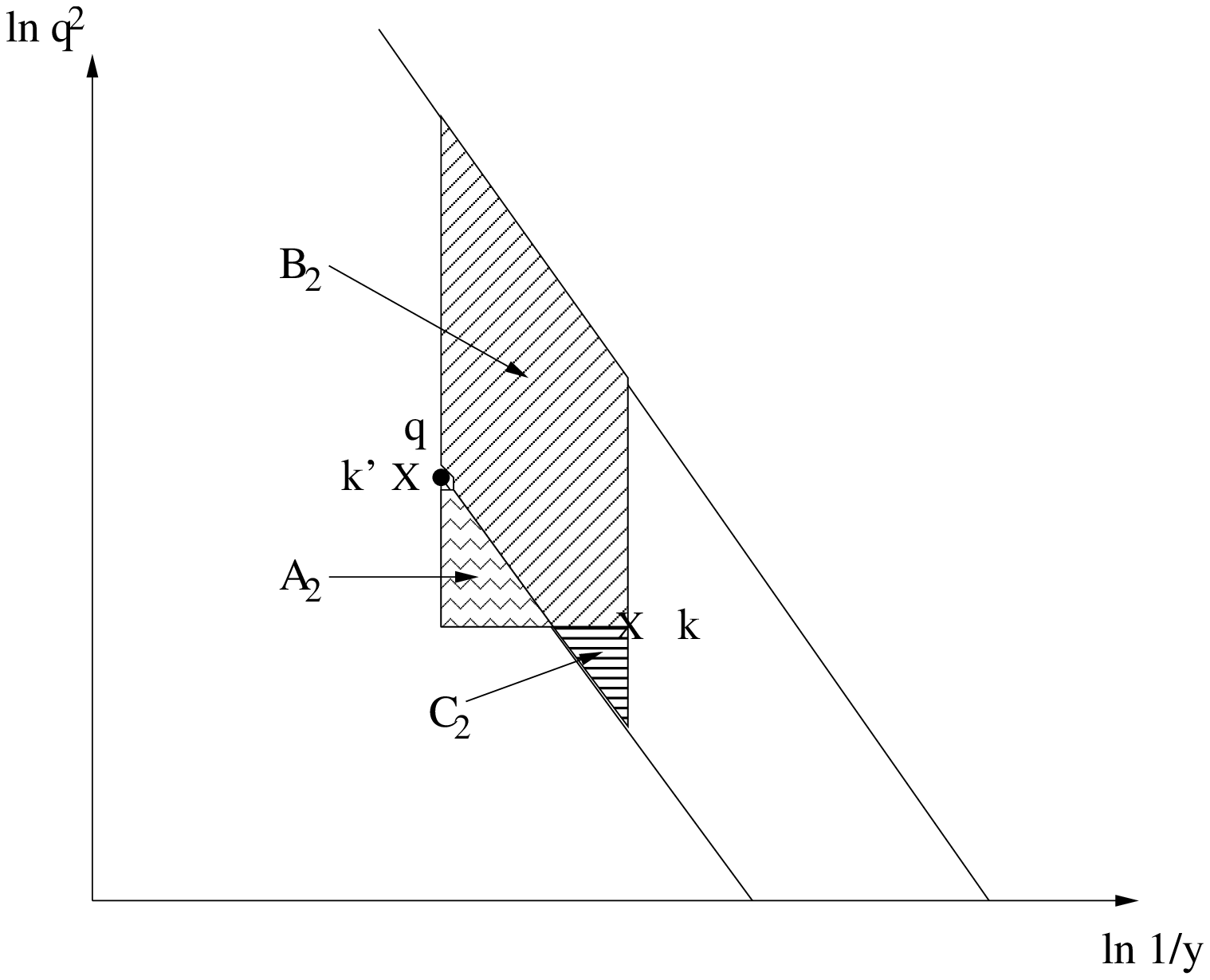}
\includegraphics[angle=0, scale=0.45]{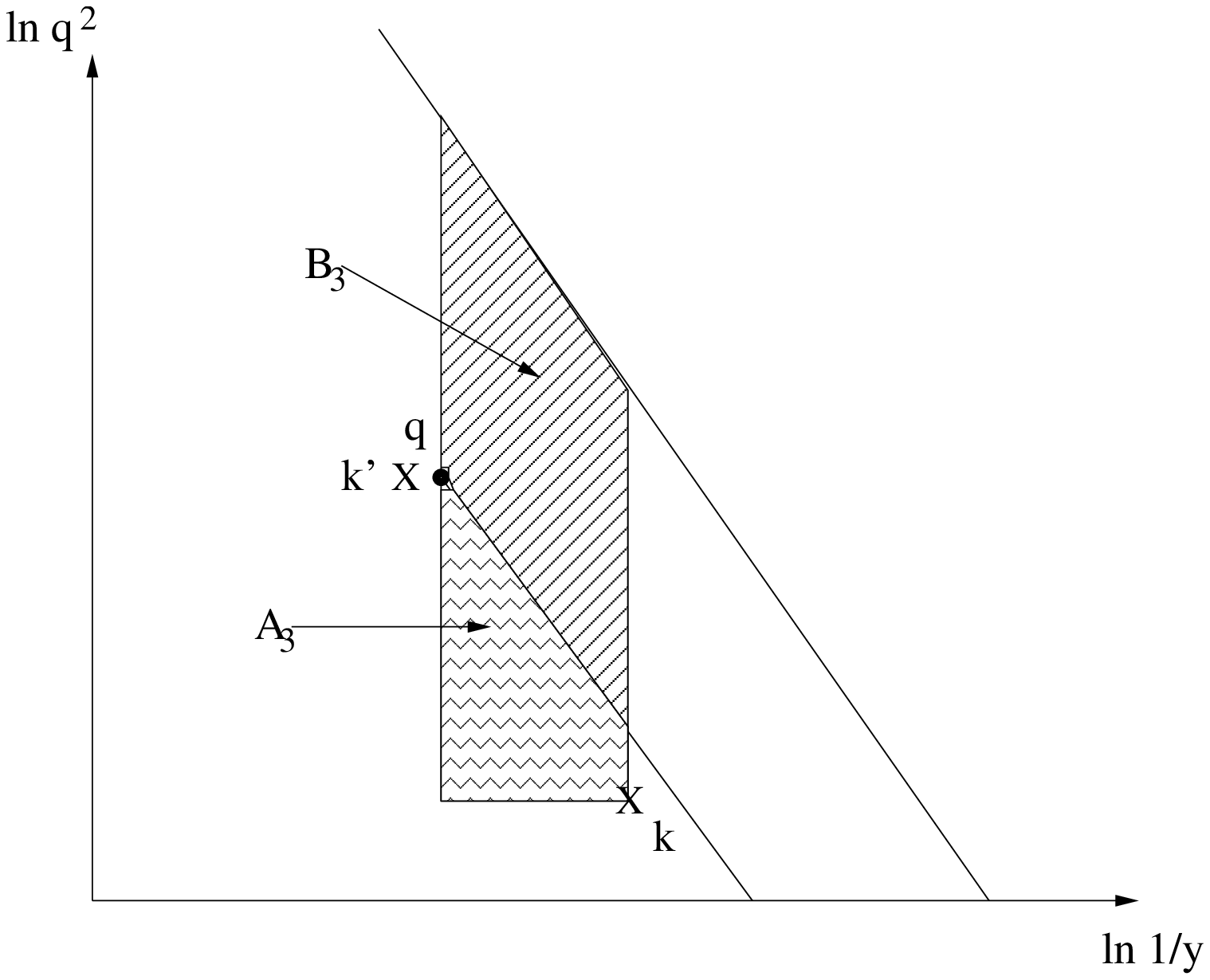}
\end{center}
\caption{\label{fig:nonSud2} Same representation as in figure \protect \ref{fig:nonSud}
but in this case with $zq \leq k < q$ (left), and $k < zq$ (right).
}
\end{figure}

As mentioned above, the non-Sudakov form factor $\Delta_{ns}$ can be written as exp$(-\abar A)$,
where in figure~\ref{fig:phasespace}, $A$ is the vertically dashed region bounded by $k$ and 
the angle of $q$, and the energies of $q$ and $k$. In CCFM, the form factors $\Delta_{ns}$ and 
$\Delta_{s}$ are derived out of the so-called eikonal, $S_{eik}$, and non-eikonal, $S_{ne}$, 
form factors which sum up to all orders the virtual corrections associated with the eikonal 
and non-eikonal emission vertices respectively  \cite{CCFM1,CCFM2,CCFM3,Marchesini:1995ia,Marchesini:1991sg}. As shown in \cite{Avsar:2009pf}, for each emission
$k$ the product of the associated form factors $S_{eik}(k)$ and $S_{ne}(k)$ equals the BFKL 
form factor $\Delta^{BFKL}(k)$, that is $S_{eik}(k) \cdot S_{ne}(k) = \Delta^{BFKL}(k)$. 
The definition of $\Delta_{ns}$ for a given emission $q$ is that it is the product 
between $S_{ne}$ and a portion of the eikonal form factor which covers a
region bounded from below by the angle of $q$ and from above by the maximal angle $\bar{\xi}$
(this is the region left over after multiplication with a soft factor coming from the 
resummed real emissions). 
In figure \ref{fig:nonSud} this is the sum of the shaded regions $B_1$ and $C_1$. The non-eikonal 
form factor $S_{ne}$ itself is such that it covers a region bounded from below by $k$ 
while from above it is again bounded by $\bar{\xi}$, this is region $B_1$ in the same figure. 
What is important however is that the sign in the exponential factor is reversed, so 
that it is positive.\footnote{This actually comes from the fact that $S_{ne}$
contains virtual corrections which are generated by the product between the non-eikonal current $J_{ne}$ 
and the eikonal current $J_{eik}$, while $S_{eik}$ is generated by the square of the 
eikonal current $J_{eik}^2$. This leads to a sign difference between the form factors, for details 
see \cite{CCFM1,CCFM2,CCFM3,Marchesini:1995ia,Marchesini:1991sg}.} The non-Sudakov is then given by 
\beq
\Delta_{ns} = \exp(-\abar(B_1+C_1)) \cdot \exp(+\abar B_1) = \exp(-\abar C_1) \; .
\label{eq:nonsud1st}
\eeq
In this case, however, we assumed that $k \geq q$, see figure \ref{fig:nonSud}. 
Now, if $k < q$ we can have two different situations, depending on whether or not $k$ 
is entirely below the diagonal line through $q$ in the figure, that is whether  
$k < zq$ or $k \geq zq$. In these both cases we define a new region $A_i$ shown in 
figure \ref{fig:nonSud2}. This is a region bounded from below by $k$ 
and from above by the angle of $q$. When $k > q$ as in figure \ref{fig:nonSud} 
of course $A_1=0$. 

On the left plot of figure \ref{fig:nonSud2} we show the case when $k \geq zq$, 
and the definition of $\Delta_{ns}$ then yields
\beq
\Delta_{ns} = \exp(-\abar(B_2+C_2))\cdot \exp(+\abar(A_2+B_2)) = \exp(-\abar(C_2-A_2))\; .
\eeq
If $A_2$ is bigger than $C_2$ then we see that $\Delta_{ns}$ is larger than unity. 
In case $k < zq$ we instead have (in this case $C_3=0$)
\beq
\Delta_{ns} = \exp(-\abar B_3) \cdot  \exp(+\abar(A_3+B_3)) = \exp(+\abar A_3) \geq 1\; .
\eeq
Thus $\Delta_{ns}$ 
exceeds unity when $A_i \geq C_i$, and this actually happens when $k^2 < z q^2$. 
All the results above give
\beq
\Delta_{ns} = \exp \left( -\abar \int_z^1 \frac{dz'}{z'} \int^{k^2}_{z'^2 q^2} \frac{dq'^2}{q'^2}
\right)  = \exp \left ( -\abar \ln \frac{1}{z} \ln\frac{k^2}{z q^2} \right)\; .
\label{eq:nonsud}
\eeq
Therefore 
if the \emph{kinematical constraint} $k^2 \geq z q^2$ is assumed to be valid, 
$\Delta_{ns}$ is guaranteed to be bounded by unity\footnote{In that case $\Delta_{ns}$, 
despite its name, works as a Sudakov and it can be used to cancel the $k_\perp$-conserving hard emissions. One is then left with a much simplified formula. This 
was first understood and used in \cite{Andersson:1995ju}, and it was used in the analysis in \cite{Avsar:2009pf}
as well. }. A different version of 
$\Delta_{ns}$ was presented in \cite{Kwiecinski:1995pu} in which case the 
regions $A_i$ are simply neglected. Then one always has 
\beq
\Delta_{ns} = \exp(-\abar C_i),
\label{JSnonsud}
\eeq
where 
\beq
C_1= \ln \frac{1}{z} \ln\frac{k^2}{z q^2}, \,\,\,\,\,  C_2=\ln^2 \frac{k}{z q}, \,\,\,\,\,  C_3 = 0 \; .
\eeq
Since $A_i$ is neglected, we are thus guaranteed that $\Delta_{ns} \leq 1$. Thus in a sense this 
has a similar effect to assuming the kinematical constraint to hold, but one should still keep
in mind that the effect is not the same because one has no apparent reason to simply 
neglect the regions $A_i$ (this amounts to neglecting soft emissions above $k$, see 
also discussion below). What the kinematical constraint ensures is that $C_i \geq A_i$.  
In this paper, we will not include the kinematical constraint (it was included in \cite{Avsar:2009pf}), which must be included 
in the real emissions as well, but we shall rather 
study the differences between using the two different non-Sudakov form factors, 
\eqref{eq:nonsud} and \eqref{JSnonsud}. 
\newline

\subsection{Scale of the running coupling}

Above we assumed a fixed coupling which is consistent with formulation of  
CCFM which was only performed at LO where the coupling is fixed. Although the effects
of the running coupling are formally of next to leading order, it is nevertheless 
for any phenomenological application extremely important to take into account 
the running of the coupling. 

The standard procedure is to let the coupling run with $k$ in the 
$1/z$ pole and in $\Delta_{ns}$ which is similar to what is usually done 
in LO BFKL. When both hard and soft emissions are included then 
$q$ is usually chosen as the relevant scale for $\abar$
in the $1/(1-z)$ pole and in $\Delta_s$ (this is for example done in CASCADE as well as in
the SMALLX program \cite{Marchesini:1990zy}). Notice that $\Delta_s$ multiplies also the $1/z$ pole so 
with this choice one is for the hard emissions using $q$ as scale for the Sudakov
form factor while $k$ is used for the non-Sudakov and the real emissions.  

As $k$ is the larger scale in $\Delta_{ns}$ it may seem natural that 
it is chosen as the scale for the running coupling. However, once a scale is 
chosen for a certain class of emissions, say the hard emissions, then that 
scale should be consistently applied everywhere because otherwise some cancelations 
which are important for the consistency of the formalism may be violated. 
Also, even though $k$ seems to be the larger scale in $\Delta_{ns}$
we have seen above that actually in the derivation 
of $\Delta_{ns}$, $k$ first enters as the lower scale in $S_{ne}$. 

In \cite{Marchesini:1990zy} it is argued that the choice to have $q$ and 
$k$ as scales simultaneously can be derived from having the single choice $(1-z)\sqrt{|k^2|}$
(where $k^2$ is now the four momentum squared), since this choice reduces 
to $k$ and $q$ when $z\to 0$ and $z\to 1$ respectively. One should, however, 
again remember that both $\Delta_{ns}$ and $\Delta_s$ are derived out of 
$S_{eik}$ and $S_{ne}$, and while the four momentum of $k$ could be argued 
as scale in the latter, the former resums the graphs where virtual gluons are emitted and 
absorbed in between the real gluons so it is hard to see why the scale 
of $k$ should also be used here. We find it more plausible that instead 
the momenta $q$ of the virtual gluons are used in $S_{eik}$. But if this 
choice is made, then to derive $\Delta_{ns}$ and $\Delta_s$ in the first place,
one has to choose $q$ as scale in $S_{ne}$ too, since otherwise necessary cancelations 
between $S_{eik}$ and $S_{ne}$ would not work. Consequently this leads to 
using $q$ in both $\Delta_{ns}$ and $\Delta_s$, and therefore also in all 
the real emissions.  

If one uses $k$ as scale in $\Delta_{ns}$ then obviously the formula 
in \eqref{eq:nonsud}, and similarly the corresponding one in \eqref{JSnonsud}
is the same as in the fixed coupling case. If on the other hand one uses 
the virtual momentum $q'$ as scale for $\abar$, then the formula is changed. 
We use $q'$ as scale in \eqref{JSnonsud} and derive a new formula which is 
however somewhat lengthy so we do not present it explicitly. We shall use 
the one loop result for the running coupling, frozen below some scale $k_0$,   
\beq
\alpha_s(q^2; k_0^2) = \frac{b_0}{\ln (\tilde{q}^2/ \Lambda_{QCD}^2)}, \,\,\,\,\,\, \tilde{q} \equiv \mathrm{max}(q,k_0)\; .
\label{eq:scaleq}
\eeq
Thus for the running coupling case we shall consider the following alternatives: One in which we use $k$ 
as scale and $\Delta_{ns}$ given in \eqref{eq:nonsud}, one in which we use $k$ as scale and $\Delta_{ns}$ in 
\eqref{JSnonsud}, and the last case where $q$ is chosen as scale in $\abar$ and the corresponding 
modified $\Delta_{ns}$ is used. 

We shall note here  that, in the BFKL equation the coupling also starts to run at the next-to-leading 
logarithmic level (in $\ln 1/x$) only \cite{Fadin:1996nw,Fadin:1997zv,Fadin:1998py,Camici:1996st,Camici:1997ij,Ciafaloni:1998gs}.  The different choices of scales formally lead to the differences 
at next-to-next-to leading level. They are however not negligible numerically. The form of the NLL correction suggests that the most natural scale 
is that of the transverse momentum squared of the emitted gluon, see for example \cite{Ciafaloni:2003rd,Ciafaloni:2003ek}.  This confirms the choice of \eqref{eq:scaleq}
as a more natural choice of scale.

\subsection{Numerical results for the linear case}
\subsubsection{Different versions of the non-Sudakov form factor}

\begin{figure}
\begin{center}
\includegraphics[angle=0, width=0.45\textwidth]{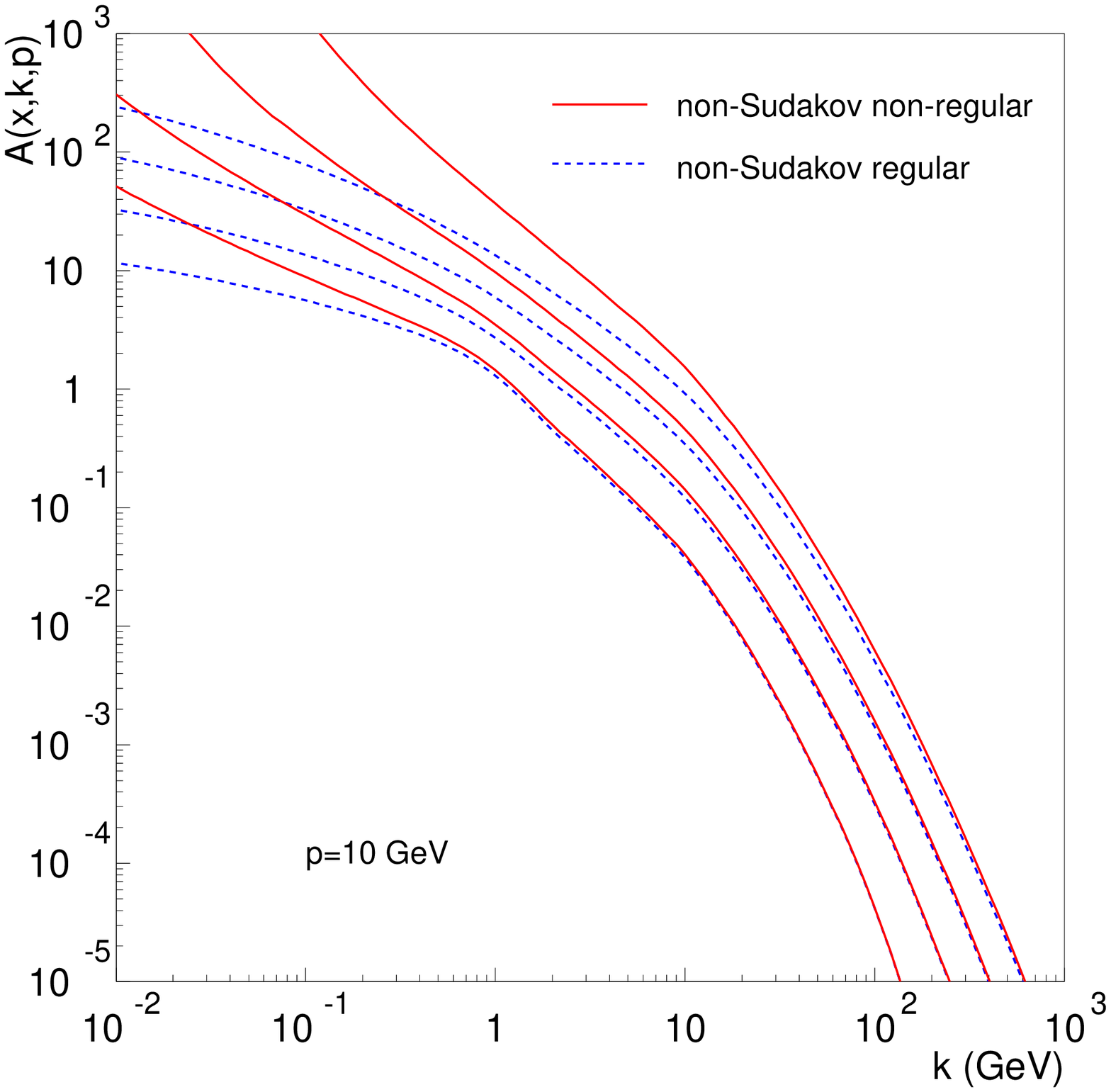}
\includegraphics[angle=0, width=0.45\textwidth]{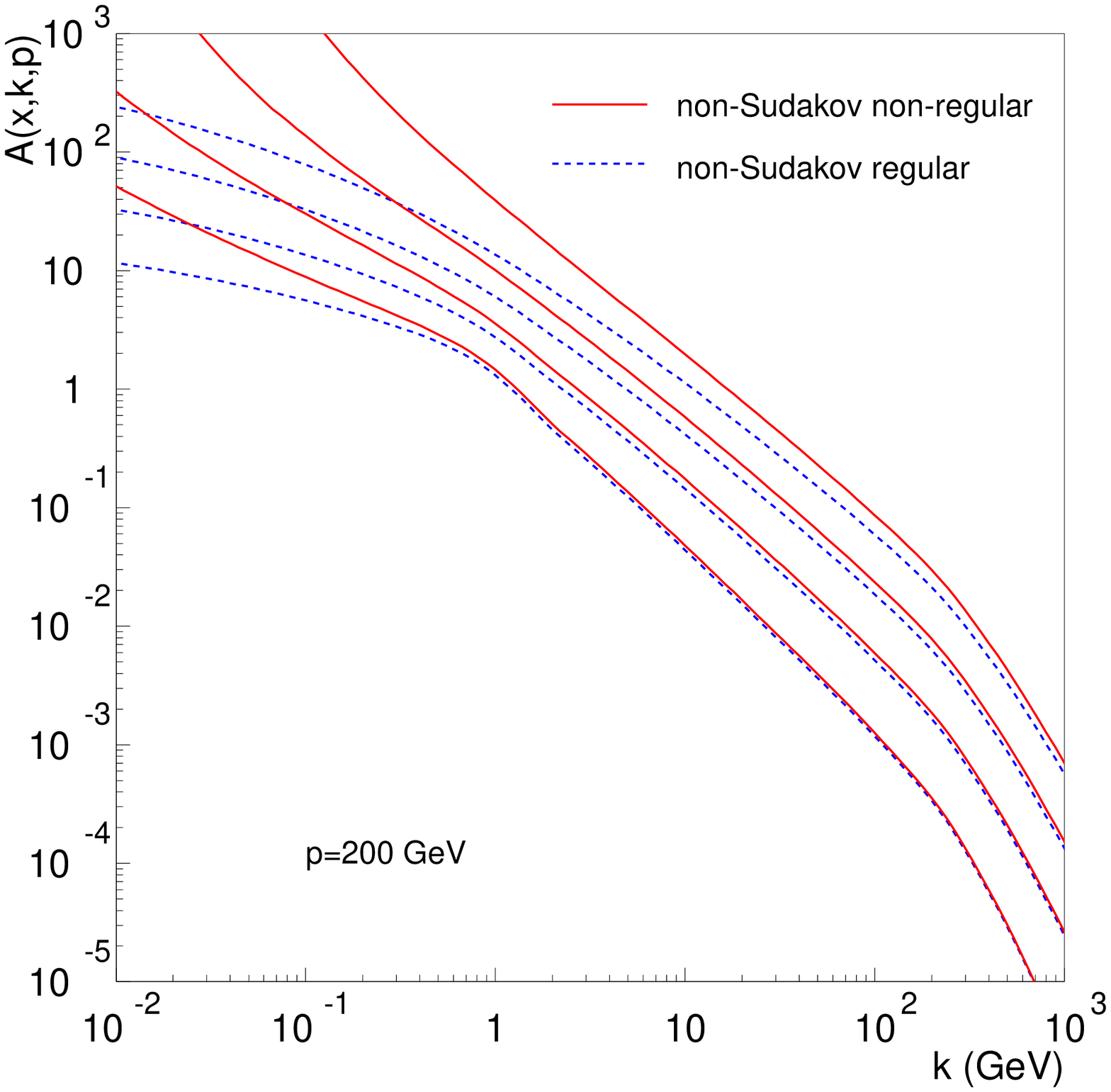}
\end{center}
\caption{Solution to the CCFM equation \protect \eqref{eq:ccfminteq2} for fixed $\abar=0.2$.  Curves increasing in magnitude are for rapidities $Y=4,6,8,10$ respectively.
Two values of the external scale are used, $p=10 \; {\rm GeV}$ (left plot) and $p=200 \; {\rm GeV}$ (right plot). Dashed blue lines correspond to the non-Sudakov form-factor (\protect \ref{JSnonsud}) while solid red curves correspond to the non-Sudakov form-factor \protect \eqref{eq:nonsud}.}
\label{fig:lindiffsud}
\end{figure}

We will present here numerical results for the solution to the linear CCFM evolution. 
 We start with the comparison of the two versions of the non-Sudakov form factor \eqref{eq:nonsud} and \eqref{JSnonsud} in the case when the coupling is fixed, and set to be equal to $\abar=0.2$. In figure~\ref{fig:lindiffsud}
we present the value of the unintegrated gluon distribution $\mathcal{A}(x,k,p)$   obtained from CCFM as a function of the transverse momentum $k$, and for the fixed value of the momentum $p=10 \;  {\rm GeV}$ (left plot) and $p=200 \;  {\rm GeV}$ (right plot). Different curves in increasing order correspond to increasing values of the rapidity. The initial condition for the equation was set to be 
\beq
\mathcal{A}^{(0)}(k,p) \; = \; \exp(-k^2/\mu^2) \, \theta(p-k) \; ,
\label{eq:initcondition}
\eeq
with $\mu=1 \; {\rm GeV}$.
It is clear that the 'non-regular' form-factor given by \eqref{eq:nonsud} gives a significant contribution in the low momentum regime due to the fact that $k^2<zq^2$ so that the expression in the exponent changes sign. This gives a substantial enhancement at low values of $k$. On the other hand it is also clear that both prescriptions for the form-factors coincide in the  large $k$ regime. One observes that the slope of the $k$ distribution becomes steeper when $k$ becomes larger than $p$, meaning a suppression of the momenta due to angular ordering.
Momenta larger than $p$ are nevertheless allowed. This change in the slope in $k$ is crucial for the analysis with the boundary and the behavior of the saturation scale as we shall see later.

\subsubsection{Fixed vs running coupling}
\begin{figure}
\begin{center}
\includegraphics[angle=0, width=0.47\textwidth]{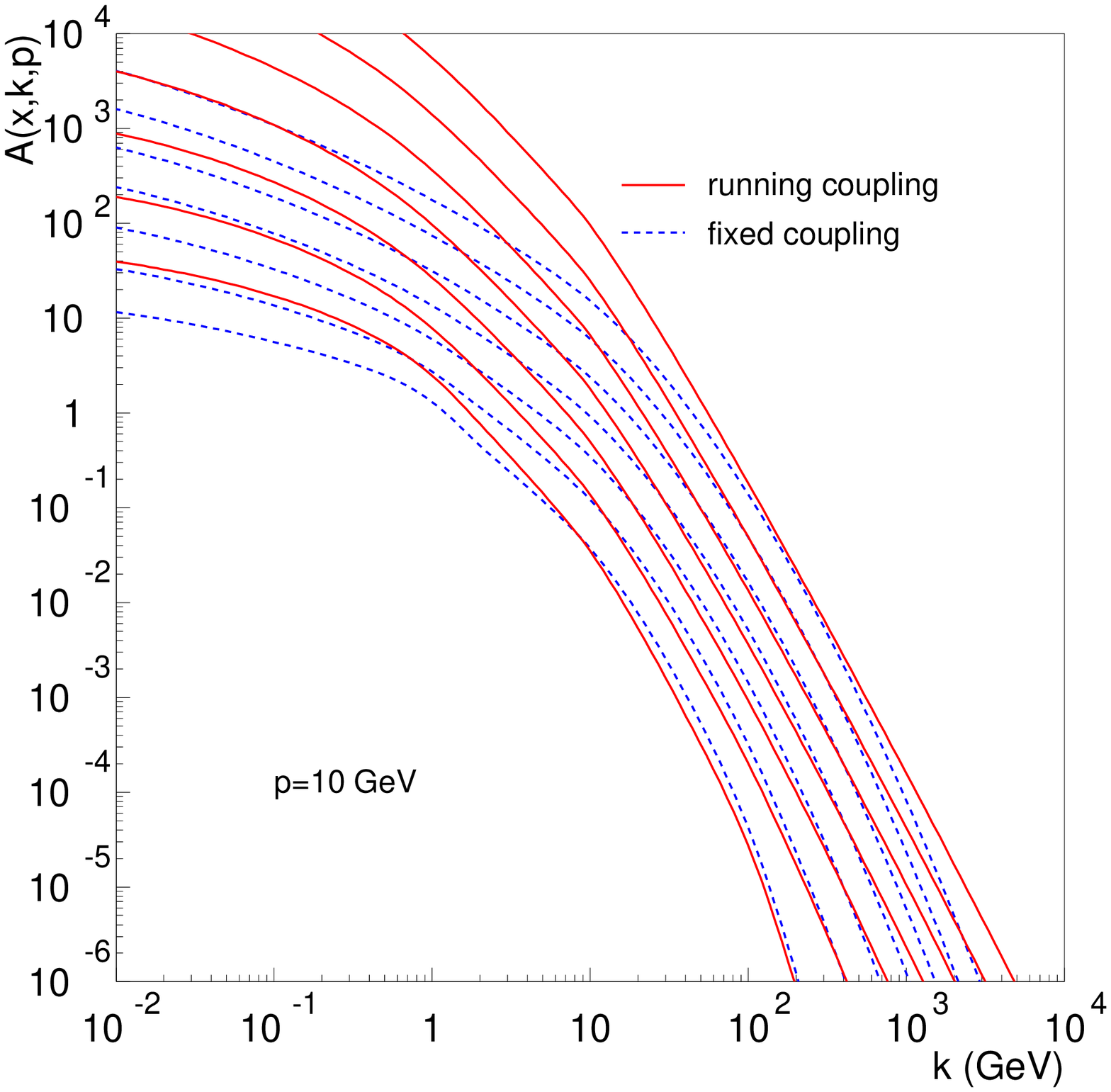}\hfill
\includegraphics[angle=0, width=0.47\textwidth]{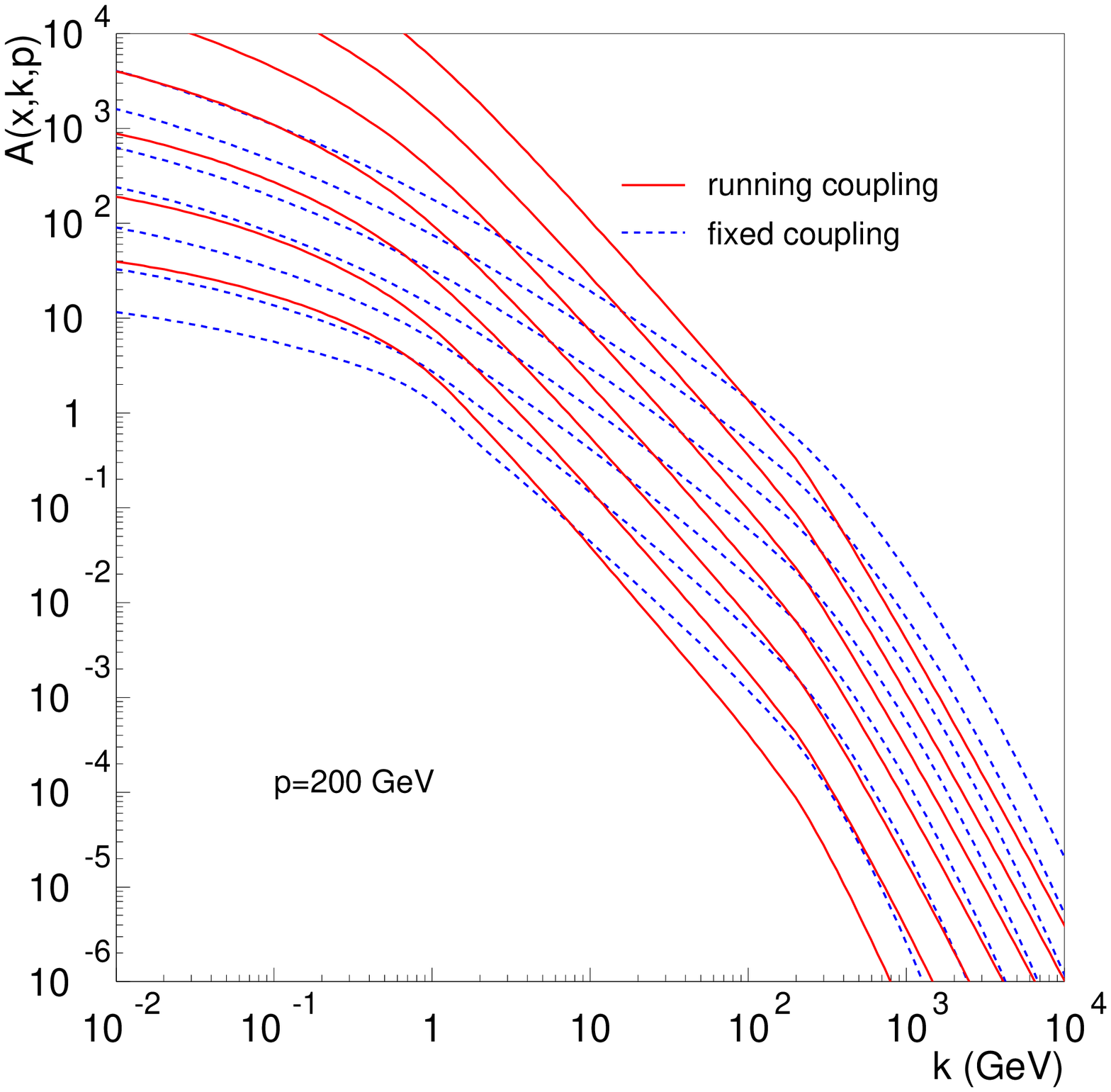}
\end{center}
\caption{Solution to the CCFM equation \protect \eqref{eq:ccfminteq2}.  Curves increasing in magnitude are for rapidities $Y=4,6,8,10,12,14,16$ respectively.
Values of the external scale are $p=10 \; {\rm GeV}$ (left plot) and $p=200 \; {\rm GeV}$ (right plot). Dashed blue lines correspond to the fixed coupling case with $\abar=0.2$ while  solid red curves correspond to the running coupling case. }
\label{fig:linfxdrun}
\end{figure}
In figure~\ref{fig:linfxdrun} we perform the comparison between the running and fixed coupling scenarios. The running coupling is regularized at $k_0^2=0.7 \; {\rm GeV}^2$ with frozen scenario. As expected the running coupling  gives the higher contribution in the region of the 
infrared momenta. It falls then below the fixed coupling at high momenta but such that $k<p$, as the running coupling value becomes smaller than the fixed value $\abar=0.2$. The effect is most pronounced for the calculation with $p=200 \; {\rm GeV}$. We observe though that in the region where $k>p$, the  scenario with the running coupling gives results which tend to be larger. One needs to take into account the fact that in this region we have stronger relative  suppression from the non-Sudakov form-factor (we shall discuss this more below). The value of the form-factor is smaller for the larger value of $\alpha_s$, which is for the case of the fixed coupling. This is why the running coupling calculation tends to be less suppressed in the asymptotic regime of the large transverse momenta.


\subsubsection{Comparison between different versions of the running coupling}
\label{sec:lindiffrun}
\begin{figure}
\begin{center}
\includegraphics[angle=0, width=0.47\textwidth]{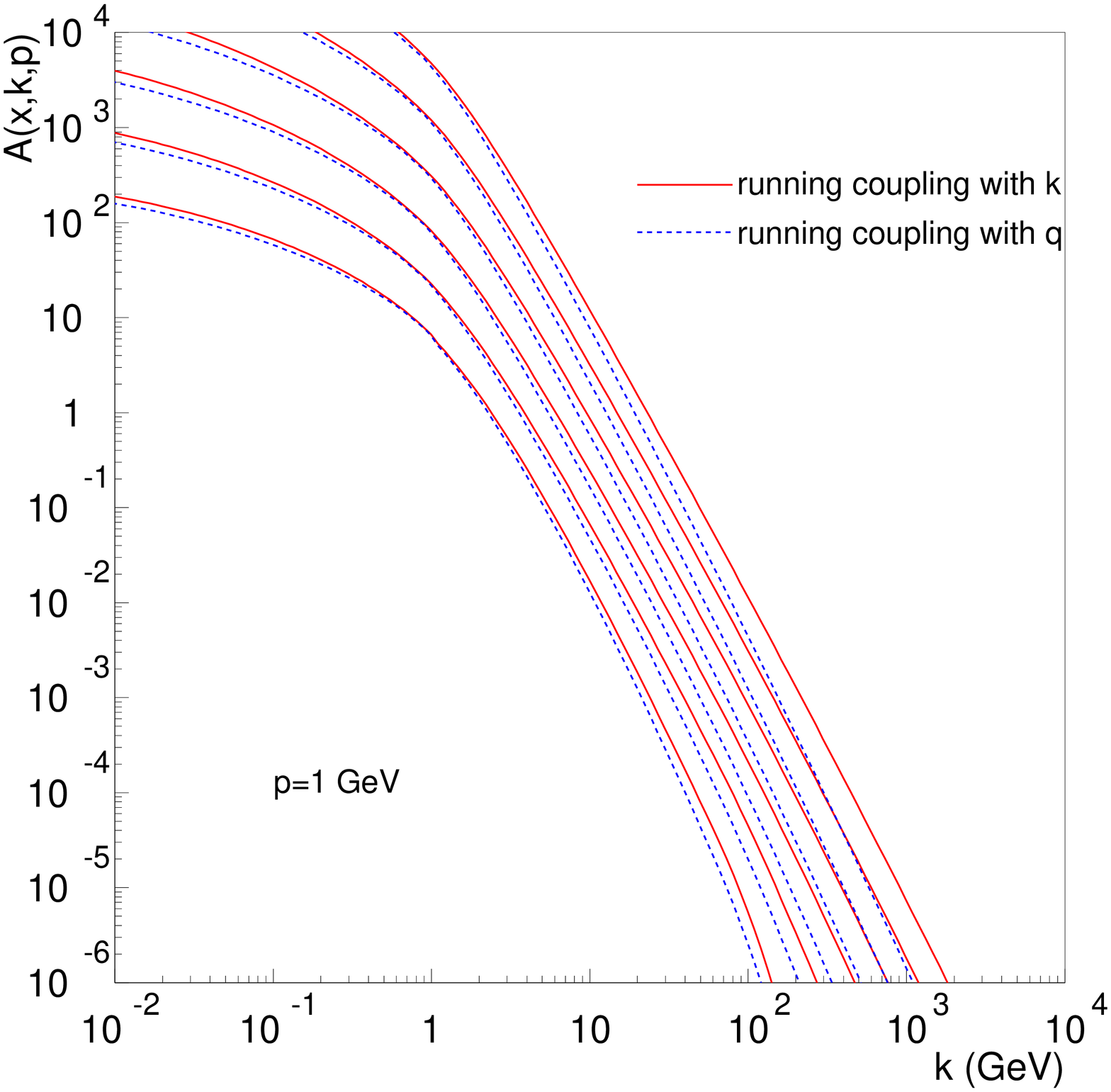}\hfill
\includegraphics[angle=0, width=0.47\textwidth]{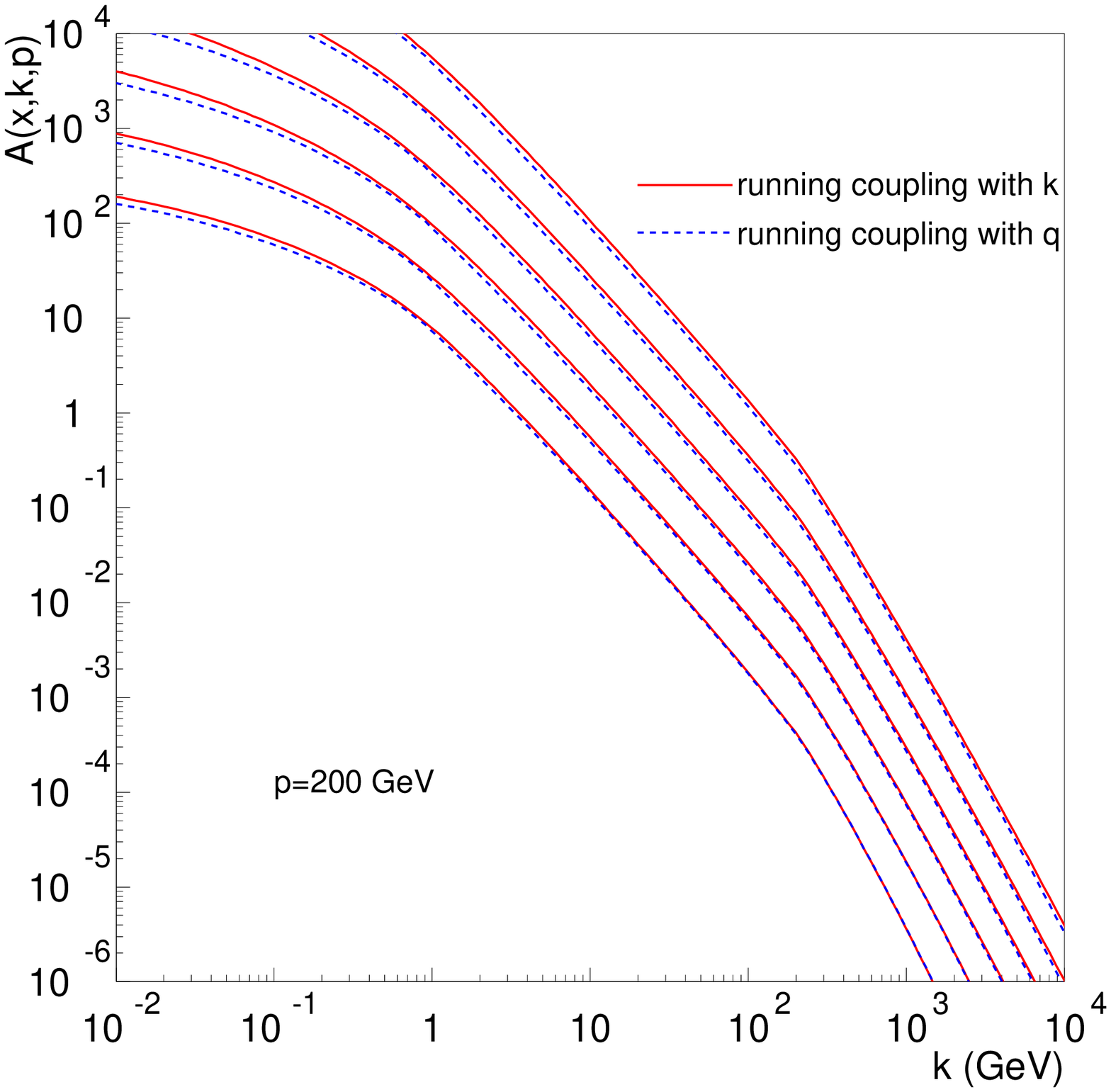}
\end{center}
\caption{Solution to the CCFM equation \protect \eqref{eq:ccfminteq2}.  Curves increasing in magnitude are for rapidities $Y=4,6,8,10,12,14,16$ respectively.
Value of the external scale $p=1 \; {\rm GeV}$ (left plot) and $p=200 \; {\rm GeV}$ (right plot). Dashed blue lines correspond to the running coupling with scale $q$; solid red curves correspond to the running coupling with scale $k$. }
\label{fig:lindrun}
\end{figure}

Next let us turn to the implementation of different scales in the running coupling. 
In figure \ref{fig:lindrun} we show the comparison between the standard results 
and those obtained using $q$ as scale.
The differences between two choices are in general not too large, with the choice of $q$ giving smaller values for the solution. In the case when $p \gg k$ there are more and more chains  with $q \gg k$ contributing,
and so typically $\abar(k) > \abar(q)$. Therefore the solution with $q$ as a choice of the scale is slightly lower.

On the other hand, when $p$ is small ($\sim 1 \;{\rm GeV}$) we see that the differences are larger, and especially
the $Y$ dependence and the slope of $\mcA$ seem to be modified. In this regime the phase space is strongly constrained by the maximum angle allowed for the 
emissions. The phase space restriction applies to the real momenta $q$, while the virtual 
momenta $k$ are not directly restricted.  Then in an event where $k \gg p$ we would expect 
that generally $k \gg q$ as well (except early in the chain). 
This implies that $\abar(k) < \abar(q)$ and one could therefore
expect to see a faster growth when $q$ is chosen as the scale. However, the opposite behavior is observed.
 This is because a larger coupling also implies that 
there is a larger suppression from $\Delta_{ns}$. We saw a hint of this behavior already in figure \ref{fig:lindiffsud}.
 When $k > p$  the 
phase space for the real emissions is more strongly constrained, on the other hand 
\eqref{eq:nonsud} or  \eqref{JSnonsud} are independent 
of $p$, so even if the real phase space is strongly constrained, the form-factor is not. 
This implies the possibility that a larger coupling in this regime actually
slows down the evolution due to the large suppression from $\Delta_{ns}$. 

\begin{figure}
\begin{center}
\includegraphics[angle=0,width=0.6\textwidth]{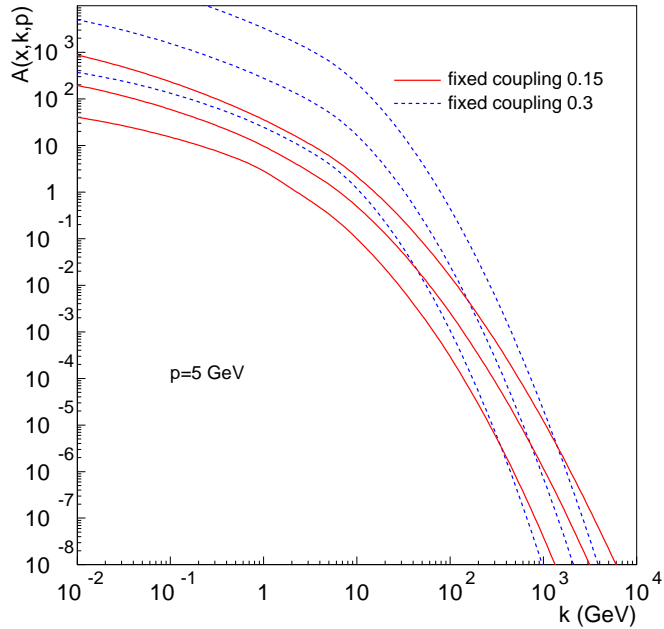}
\end{center}
\caption{Solution to the CCFM equation \protect \eqref{eq:ccfminteq2}.  Curves increasing in magnitude are for rapidities $Y=8,12,16$ respectively.
The value of the external scale is $p=5 \; {\rm GeV}$. Dashed blue lines correspond to $\abar=0.3$ while the solid red curves correspond to $\abar=0.15$. }
\label{fig:linfxdfxd}
\end{figure}

To further demonstrate that this is indeed the case we show in figure \ref{fig:linfxdfxd} results
obtained from two different values of the fixed coupling, $\abar = 0.15$ and $\abar =0.3$, for $p=5$ GeV. For clarity of demonstration  we show only solutions for $Y=8, 12, 16$.
Of course in this case since 
$\abar =0.3 > \abar = 0.15$ for any $k$, in the region where $k \lesssim p$ 
the solution with the larger coupling indeed grows faster (the phase space restriction 
is not so important here). However, we then clearly 
see that as $k > p$, the suppression from $\Delta_{ns}$, which is much larger 
when $\abar =0.3$, starts to dominate and therefore the solution with the larger
coupling is strongly suppressed. Consequently in the region $k \gg p$, 
the solution with the larger coupling actually grows slower.

\section{CCFM in the presence of saturation: analytical results}

Having reviewed some of the basic concepts in CCFM we now move on to the question 
of implementing saturation effects and finding the relevant saturation scale $Q_s$. 
The implementation of saturation effects will be done using the same approach as 
in \cite{Avsar:2009pv, Avsar:2009pf}, but most importantly because in this 
work we do keep the full dependence of $\mcA$ on the additional scale $p$, 
we shall discover many interesting new results not noticed before. The main 
implementation is numerical and all the numerical results are to be presented 
in section \ref{sec:results}. In this section we will analytically study 
the general behavior of $Q_s$. 

To study the saturation scale it is convenient to take the Mellin transform over the momentum $k$. 
Following \cite{Bottazzi:1997iu} we shall write the Mellin representation 
of $\mcA$ as 
\beq
\mcA(Y,k,p) = \melint \, e^{\omega(\gamma)Y-(1-\gamma)\rho}\, H(p/k) \; .
\label{Mellin}
\eeq
where $Y = \ln 1/x$, and we define $\rho = \ln (k^2/Q_0^2)$ with 
$Q_0$ an arbitrary scale. We will only consider the fixed coupling case in the 
analytic formulas. The function $H$ encodes the dependence of $\mcA$ on the 
parameter $p$. What we know generally is that $H \approx 1$ when $p/k \gg 1$, 
since the restriction coming from $\bar{\xi}$ becomes irrelevant in that 
case. This, however, does not imply that angular ordering is not important 
anymore. More precisely it means  that the ordering  in the final step of the evolution 
is automatic when $p \gg k$, but  in the previous steps of the evolution
it is still important. This will reflect itself in the characteristic 
function $\omega$ which in the limit $p \gg k$ is given by
\beq
\omega = \abar \qint \left( \left(\frac{k^2}{\kprime^2}\right)^{1-\gamma}H(q/\kprime)
- \theta(k-q)\, H(q/k) \right).
\label{ccfmomega}
\eeq

To find the function $H$ one can differentiate the integral equation \eqref{eq:ccfminteq2}
with respect to $p$ as done in \cite{Bottazzi:1997iu}. One then finds that 
\beq
\frac{\partial H(p/k)}{\partial \ln p} = \abar \qint \, \theta(Y - \ln q/p) \, \theta (q - p)
\, \Delta(\ln q/p, k, q) \, \left(\frac{p}{q} \right)^{\omega}\! \left (
\frac{k^2}{k'^2}\right)^{1-\gamma}\!\!\! H(q/k') \; ,\nonumber \\
\label{Hequation}
\eeq
where as before $k'=\kprime$.
Assuming that $Y$ is very large and the solution is in the asymptotic region one can  neglect 
the first theta function, which is what is done in \cite{Bottazzi:1997iu}.
However, 
one should then notice that with the non-Sudakov in \eqref{eq:nonsud} which is allowed to 
exceed unity, one would run into a divergence. This is so 
because when $Y\to \infty$, even if $p$ is finite, the phase 
space for the emissions contributing to the scattering becomes infinitely large, 
and there is  a UV divergence appearing when $q \to \infty$.
On the other hand such a divergence does not appear if the restricted 
$\Delta_{ns}$ in \eqref{JSnonsud} is used instead. The reason for this is 
that the choice  in \eqref{JSnonsud} effectively implies that $k$ plays the role of a
UV cut (but only for the soft emissions). If the restriction from $\bar{\xi}$ is removed (as when $Y \to \infty$)
then the form factors $S_{eik}$ and $S_{ne}$ need to be regularized by some 
UV cut-off. When these factors are multiplied together, because of the different 
signs, they cancel each other in the region above $k$, and thus the dependence 
on the UV cut-off vanishes. However, the real emissions need still be regularized 
by the UV cut-off, since otherwise the inclusive summation over the soft 
emissions for the given hard emission $q$ would give a divergence as $q\to \infty$. 
The fact that the regions $A_i$ in figures \ref{fig:nonSud} and \ref{fig:nonSud2} 
are neglected in the derivation of \eqref{JSnonsud} means that one is cutting 
off all real soft emissions above $k$ which is why $k$ plays the role of the UV 
regulator\footnote{At this point it should, however, be mentioned that there is no reason 
for the soft emissions to be bounded by $k$. This would essentially be true if the soft emissions 
were $k_\perp$ conserving in the angular ordered cascade. The fact is that they are $k_\perp$ conserving,
but \emph{not} with respect to the angular ordering, but rather to the \emph{energy ordering} 
whereby each emission is indexed according to its energy, and the ''propagator'' momenta are 
defined accordingly.} for the soft emissions. 
Thus for simplicity we shall use this form factor in the following, 
though a proper renormalization procedure for all the emissions would probably be called for.

After these remarks, let us now return to the equation for $H$ above. 
The saturation of the solution can be analyzed in the two different regimes of large and small $p$. 
\newline

\emph{The case when $p > k$:}
\\

 In case $p > k$,
since $q > p$ (because of the theta function in \eqref{Hequation}), 
we also have that $q > k$. We will then approximate $k' \approx q$. 
Equation \eqref{Hequation} is then easily solved to give 
\beq
H(p/k) &=& 1 - \frac{\abar\, H(1)}{(1-\gamma)(\omega +2 - 2\gamma)}
(1 - e^{-(\omega +2 - 2\gamma)Y})\left( \frac{k^2}{p^2}\right)^{1-\gamma} \nonumber \\
&=& 1 -  \frac{\abar \, H(1)}{(1-\gamma)(\omega +2 - 2\gamma)}
e^{-(1-\gamma)(\rho_p-\rho)} \; ,
\label{Hsolution1}
\eeq
where in the second equality we neglected the term exponential in $Y$,  because it is very small when $Y$ is large.
We have also  defined $\rho_p \equiv \ln (p^2/Q_0^2)$. Consistency of the solution requires that 
\beq
H(1) = \frac{1}{1+\frac{\abar}{f(\gamma)}}, \,\,\, \mathrm{where} \,\,\,\, f(\gamma) \equiv (1-\gamma)(\omega +2 - 2\gamma) \; .
\eeq
We see that when $(\rho_p-\rho) \gg 1$, that is when $p$ is very large, 
$H$ reduces to unity.  Since $(\rho_p-\rho)$ is large we can rewrite $H$ 
as 
\beq
H(p/k) = \exp \left (- \frac{\abar \, H(1)}{f(\gamma)}e^{-(1-\gamma)(\rho_p-\rho)}\right )\; .
\eeq
The Mellin representation then reads
\beq
\mcA(Y,k,p) \approx \melint \,e^{\omega Y - (1-\gamma)\rho -\frac{\abar H(1)}{f(\gamma)} e^{
-(1-\gamma)(\rho_p - \rho)}} \; .
\label{Mellinrep1}
\eeq 

To find the saturation scale, one can essentially follow a line of constant 
density \cite{Mueller:2002zm, Iancu:2002tr} which means that the relevant 
saddle point for the saturation problem is determined by the equations (defining 
$\rho_s \equiv \ln (Q_s^2/Q_0^2)$)
\beq
\omega_s Y &-& (1-\gamma_s)\rho_s -\frac{\abar H(1)}{f_s}\,  e^{-(1-\gamma_s)
(\rho_p - \rho_s)}= 0 \; , \label{ccfmsaddle1}\\
\omega_s' Y + \rho_s\!\! &-& \!\!\frac{\abar H(1)}{f_s}e^{-(1-\gamma_s)
(\rho_p - \rho_s)}\left(\frac{H'(1)}{H(1)}-\frac{f'_s}{f_s}
+ \rho_p - \rho_s \right)
 = 0 \; .
  \label{ccfmsaddle2}
\eeq
Here all derivatives are with respect to $\gamma$ and 
the second equation is the usual saddle point condition while the first 
equation states that the integrand around the saturation scale is constant and of 
order 1. On the other hand we recall that in the saturation problem for BFKL
the corresponding saddle point equations read
 \beq
\omega_s Y - (1-\gamma_s)\rho_s = 0 \label{bfklsaddle1}, \\
\omega_s' Y + \rho_s = 0.\label{bfklsaddle2}
\eeq

What we notice, besides the fact that the equations in CCFM look much more complicated 
than in BFKL, is that equations \eqref{ccfmsaddle1} and \eqref{ccfmsaddle2} imply 
that the 
saturation anomalous dimension $\gamma_s$, and thus also the characteristic function 
$\omega_s$, depend generally on $p$ and $Y$. In BFKL on the other hand, $\gamma_s$ is a pure number and 
$\omega_s$ is independent of $Y$ which we see from the corresponding equations. 
As $\rho_p - \rho_s \to \infty$, we see that the structure of the CCFM saddle point equations 
reduce to the BFKL ones, though at finite $\abar$, $\gamma_s$ and $\omega_s$ are 
still different than in BFKL. When $\abar \to 0$, the CCFM solution reduces completely 
to the BFKL one. 

From equation \eqref{ccfmsaddle1} we can find $\rho_s$. Let $\rho_s^{(0)}$ 
be the solution when $\rho_p - \rho_s \to \infty$, that is 
\beq
\rho_s^{(0)} = \frac{\omega_s^{(0)}}{1-\gamma_s^{(0)}}\,Y \; .
\eeq
 Since the general form of $\omega$ which depends on $p$ is complicated, 
 we will in the following simply make the approximations to set $\gamma_s=\gamma_s^{(0)}$
 and $\omega_s = \omega_s^{(0)}$. Then let us write\footnote{The minus sign in front of the correction term comes from the fact that a finite $p$ 
reduces the phase space and should therefore reduce $\rho_s$ too.} $\rho_s=\rho_s^{(0)}-\rho_s^{(1)}$, 
and we get 
 \beq
\tilde{\rho}_s - \frac{\abar H(1)}{f_s}\,e^{-(1-\gamma_s)
(\rho_p - \rho_s)} \, e^{-\tilde{\rho}_s}=0, \,\,\, \mathrm{where} \,\,\,  \tilde{\rho}_s \equiv (1-\gamma_s)\,\rho_s^{(1)} \; .
\eeq
Since $(\abar H/f)e^{-(1-\gamma_s)(\rho_p - \rho_s)} \ll 1$ (it is suppressed by both 
$\abar$ and the exponential) we must have $\tilde{\rho_s}  \ll 1$, and therefore 
\beq
\tilde{\rho}_s \approx \frac{\abar H(1)}{f_s}e^{-(1-\gamma_s)
(\rho_p - \rho_s)}\left( 1 - \frac{\abar H(1)}{f_s}e^{-(1-\gamma_s)
(\rho_p - \rho_s)}\right) \; .
\eeq
Thus we find that 
\beq
\rho_s = \frac{\omega_s^{(0)}}{1-\gamma_s^{(0)}}\, Y- 
\frac{\abar H(1)}{(1-\gamma_s^{(0)})f_s^{(0)}}e^{-(1-\gamma_s^{(0)})
\rho_p + \omega_s^{(0)}Y} + \cdots \; ,
\label{eq:satscale1}
\eeq
which is valid as long as $\rho_p - \rho_s^{(0)} \gg 1$.
The saturation saddle point is determined by 
\beq
-\omega^{(0)\prime}_s = \frac{\omega_s^{(0)}}{1-\gamma_s^{(0)}} \; , 
\label{saddleeq}
\eeq
as in BFKL. Of course this is the lowest order approximation (lowest order with respect 
to the corrections in $p$, but it is valid at finite $\abar$) but since the 
more general case is rather complicated we shall use this formula. 
In the later section we will determine the full saturation momentum via the numerical solution.

\begin{figure}
\begin{center}
\includegraphics[angle=0, width=0.6\textwidth]{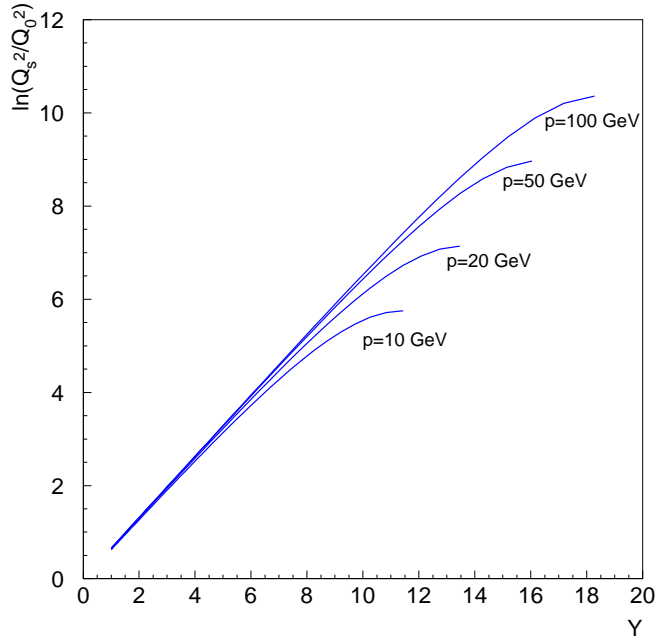}
\end{center}
\caption{Logarithm of the saturation scale from (\protect \ref{eq:satscale1}) as a function of rapidity. The curves in increasing order are for 
external scales \protect $p=10,20,50,100\;{\rm GeV}$. The scale $Q_0=1 \; {\rm GeV}$.
The curves are truncated at the limit of the applicability of the approximation used.}
\label{fig:satscalemodel}
\end{figure}

In order to be able to use the formula (\ref{eq:satscale1}) one needs to have a solution 
for the characteristic function in CCFM. This is essential for obtaining $\omega_s^{(0)}$ 
and $\gamma_s^{(0)}$.  Unfortunately it is only possible to extract the characteristic function 
numerically, as was done in \cite{Bottazzi:1997iu}. To estimate here the dependence of the 
saturation scale on $p$ and $Y$ from (\ref{eq:satscale1}) we will use a simple model 
for the characteristic function,
\beq
\chi(\gamma,\omega) = 2 \Psi(1) - \Psi(\gamma) - \Psi(1-\gamma+\omega) \; . 
\eeq
This model corresponds to the LO BFKL equation with the kinematical constraint included. It 
is well known that it leads to the significantly reduced value of the intercept \cite{Kwiecinski:1997ee,Salam:1998tj} and thus it 
should be fairly close numerically to the CCFM characteristic function. One can extract 
$\gamma_s^{(0)}$ and $\omega_s^{(0)}$ solving numerically equation (\ref{saddleeq}).  For the 
standard  choice of the coupling constant $\abar=0.2$ we obtained $\gamma_s^{(0)}=0.44$,  
$\omega_s^{(0)}=0.37$ and $-\omega_s^{'(0)}=0.66$.
Using these extracted values we can determine saturation scale  from  the 
equation (\ref{eq:satscale1}). We demonstrate the results in figure~\ref{fig:satscalemodel}.
The  curves increasing are for different values of the momentum scale $p=10,20,50,100 \;{\rm GeV}$.

It is evident that for small rapidities the growth of the saturation scale is dominated by the term exponential in rapidity and the characteristics is similar  to the BK equation (or BFKL with the saturation boundary).
However for fixed value of scale $p$ the dependence of $\rho_s$ on rapidity becomes nonlinear. The smaller the value of $p$ is, the lower is the critical value of rapidity for which the behavior starts to become nonlinear, and there is an additional dependence on $p$.
The formula cannot be trusted in the region where $\rho_s$ falls off with rapidity. This region is outside the validity of the original approximation in which equation~(\ref{eq:satscale1}) was derived. 
One would need to compute higher order corrections  and possibly resum them in order to predict the asymptotic behavior of $Q_s$ with $Y$.

Thus the leading term for the saturation momentum when $p$ is very large has the same 
structure as in BFKL, with the ''speed'', $-\omega_s'$,  of $\rho_s$ being 
much slower, around $0.65$ for CCFM as opposed to around $0.98$ for BFKL 
at LO and $\abar =0.2$. The corrections coming from finite $p$ slow 
down the saturation scale and from \eqref{eq:satscale1} we see that 
the $p$ dependence in $Q_s^2$ enters as a power inside an exponential.
 Summarizing,  
we observe that the corrections from the phase space restriction in CCFM 
introduce an increasingly complicated shape of $Q_s$ depending both on $p$ and $Y$ 
when going beyond the BFKL regime, i.e. $p\gg k$.

Using the results above one can also study the behavior of $\mcA$ near saturation, 
that is when $0 < \rho - \rho_s \lesssim 1$. 
This is done by expanding the function in the 
exponential of the Mellin representation to $\mcal{O}((\rho - \rho_s)^2)$. In 
BFKL this leads to
\beq
\mcA_{BFKL} \sim \frac{1}{\sqrt{2\pi\omega_s''Y}}e^{-(1-\gamma_s)(\rho-\rho_s)}
e^{-\frac{(\rho-\rho_s)^2}{2\omega_s''Y}} \; . 
\label{BFKLfunc}
\eeq
The BFKL result features the well known 
characteristics of the small-$x$ evolution. One is the property of geometric 
scaling which is the fact that the $Y$ and $k_\perp$ dependence enters 
the gluon distribution in the combination $k_\perp/Q_s$, that is via $\rho-\rho_s$.
Indeed as long as $\rho-\rho_s \ll \sqrt{2\omega_s''Y}$ the dominant dependence 
of $\mcA$ is on $\rho-\rho_s$. A more careful treatment performed in \cite{Mueller:2002zm}
modifies this solution so that the $Y$ dependent pre-factor coming from the 
Gaussian integration is replaced by a factor $(\rho-\rho_s + \Delta)$ for some 
number $\Delta$. In this case geometric scaling is even more obvious. 
The second property is the diffusion in $k_\perp$ space. 
This is governed by the Gaussian and the diffusion radius grows as $\sqrt{2\omega_s''Y}$. 

One can perform the exact same manipulations also in \eqref{Mellinrep1}. 
However the resulting expression is rather long and not very transparent 
so we do not write it down here. Again in case $\rho_p - \rho \gg 1$ 
the structure reduces to that in  \eqref{BFKLfunc}. If the $p$ dependence 
is kept, both of the two characteristics mentioned above are modified. 
For example, diffusion is reduced due to angular ordering, and 
the diffusion radius obtains a complicated dependence on $\rho_p - \rho$. Geometric
scaling is also not transparent anymore, because of the additional 
$p$ and $Y$ dependence. Besides, generally we would need to take into 
account the $p$ and $Y$ dependence of $\gamma_s$ and $\omega_s$ which 
further breaks the scaling behavior. One may find a modified scaling behavior
even if it is not as transparent as in the BFKL case, but given the complexity 
of the solutions we shall refrain from pursuing this question analytically.  
\newline

\emph{The case when $p < k$:}
\\

\comment{Let us now turn to the opposite case when $k > p$. In this situation, 
equation \eqref{ccfmomega} is no longer valid for $\omega$ since we there assumed
that $p$ was much larger than $k$. Since the general expression for $\omega$ 
is rather complicated we shall not write it down here. The equation for $H$ 
is still the same however. In this case, the $q$ integral can be divided 
into the regions where $q < k$ and 
$q > k$. In the first region approximately $k' \approx k$ while in the second region $q \approx k'$. 
Using these crude approximations one can perform the integrals easily and 
arrive at the solution
\beq
H(p/k) = \frac{2\abar C}{\omega} e^{- \frac{\omega}{2}(\rho-\rho_p)} e^{-\abar (\rho - \rho_p)^2}
\eeq
where $C$ is a function of $\abar$ and $\omega$, but is independent
of $p$ and $k$. In this case the saddle point equations read 
\beq
\omega_s Y - (1-\gamma_s)\rho_s - \frac{\omega_s}{2} (\rho_s - \rho_p) - 
\frac{\abar}{4}(\rho_s-\rho_p)^2 = 0 \label{ccfmsaddle3} \\
\omega_s' Y + \rho_s - \frac{\omega'_s}{2} (\rho_s-\rho_p) = 0 \; .
 \label{ccfmsaddle4}
\eeq
In this case one can write 
\beq
Q_s^2(p,Y) = Q_0^2\left( \frac{p}{Q_0} \right)^{\Gamma_s(p,Y)}
e^{\Gamma_s(p,Y) Y} \; ,
\eeq
where 
\beq
\Gamma_s(p,Y) \equiv \frac{2\omega'_s(p,Y)}{\omega'_s(p,Y) -2 } \; .
\eeq
As before we can also expand the integrand close to the saturation scale 
and then perform the Gaussian integrals to get}

The opposite region where $p \ll k$ is more complicated, but also turns out to be 
much more interesting. In this case we can no 
longer use expression \eqref{ccfmomega} for $\omega$, and the $p$ dependence
is stronger. Note that in the argument above, 
which follows the one in \cite{Iancu:2002tr}, saturation is not 
really explicit. It is understood that the solution represented by the 
Mellin transform is valid only as long as one is above $Q_s$, but the effects
of saturation are kept implicit. Thus one could actually get the same 
result if one simply took linear BFKL and followed a line of constant 
amplitude, though of course in the linear case this is rather ad hoc since there 
is nothing special about any particular line of constant amplitude, nevertheless the answer
for $Q_s$ would be the same. Therefore the leading asymptotic $Y$ behavior of $Q_s$ extracted in this
way from BFKL, without actually explicitly applying saturation, is the same 
as that extracted from the non-linear BK equation. To obtain correctly the 
non-leading $Y$ behavior, however, one has to go beyond this approach 
but what was shown in \cite{Mueller:2002zm} is that one can get the correct
answer from the linear equation if the effects of saturation are 
 introduced via the boundary prescription. 

In CCFM the effects of saturation will be seen to be rather dramatic.  
From the numerical analysis we shall see that the phase space restriction 
coming from $p$, which has its origin in the coherence of the QCD radiation, 
when combined with saturation, which imposes a restriction 
in the region below $Q_s$, 
implies that the saturation scale itself saturates at a region $Q_s \gg p$. This 
behavior is missed by the naive analysis above since one must 
explicitly apply saturation to possibly see it. A treatment like in \cite{Mueller:2002zm}
may lead to an analytic understanding of this novel behavior, but given the 
complexity of the equations involved this may be very difficult to pursue. 
We shall therefore not apply the analytic treatment to this case, 
even though finding $H$ itself is easy, as done in \cite{Bottazzi:1997iu}. 
Instead we move on to the numerical treatment.

\section{Description of the boundary method}
\label{sec:numerics}

The boundary method applied to the linear BFKL evolution, was shown in \cite{Avsar:2009pv, Avsar:2009pf} to successfully 
reproduce the results obtained from the non-linear BK equation.  The saturation 
momentum $Q_s$ and the shape of the gluon distribution for $k \geq Q_s$ were correctly reproduced by this method. The 
basic idea is to prevent the gluon occupation number from 
growing too large. First, we define a line of constant 
occupation number via the condition
\beq
\mcA(Y, \rho_c, p) = c \; ,
\eeq
where the number $c$ is of order 1 or smaller (but not \emph{much} smaller). As obvious from 
the definition, $\rho_c$ depends on \emph{both} $Y$ and $p$. In the linear evolution, 
$\mcA$ will continue to grow indefinitely beyond $\rho_c$ when increasing $Y$. We then enforce a boundary condition preventing such an unrestricted growth beyond 
the critical line $\rho_c$. In practice this is done by applying the boundary for all 
momenta $\rho \leq \rho_c(Y,p) - \Delta$. Here $\Delta$ is 
a given number. Strictly speaking in CCFM $\Delta$ should not be a pure number but should 
depend on $Y$ and $p$. However, given that explicit analytic solutions of CCFM are not 
available, it is not really feasible to analytically find the possible dependence of 
$\Delta$ on $Y$ and $p$. We will therefore for simplicity let $\Delta$ be a constant 
as in BFKL. The parameters $c$ and $\Delta$ can be viewed as free parameters although in 
BFKL it is known that $\Delta \sim \ln(1/c)$. In particular the results should not depend 
on the precise values of these parameters, apart from differences in normalization. 
Our default choice will be to set $\Delta =2$ and $c=0.4$. 
In \cite{Avsar:2009pv, Avsar:2009pf} the default choice of the boundary condition was to 
let 
\beq
\mcA(Y, \rho, p) = 0, \,\,\,\, \mathrm{for} \,\, \mathrm{all} \,\,\,  \rho \leq \bar{\rho}_c(Y,p)
\equiv \rho_c(Y,p) - \Delta \; .
\label{boundary0}
\eeq
In this case one has a totally absorptive boundary, and this was indeed 
also the choice in the analytical analysis in \cite{Mueller:2002zm}.
This condition is convenient for the analytical analysis in the BFKL case because 
 then the problem can be formulated as 
diffusion in the presence of an absorbing wall \cite{Mueller:2002zm}.

 The precise 
choice of the boundary condition, should (at least asymptotically) not make any difference 
beyond a normalization factor. Indeed ,
it was explicitly demonstrated in  \cite{Avsar:2009pv, Avsar:2009pf} using the numerical 
analysis that the method is robust for any $Y$ (thus not only asymptotic ones), so that different choices still give similar 
results. In this paper we shall not be using the above condition, but we will rather 
investigate different alternatives. 

One possible choice of boundary is to let $\mcA$ grow logarithmically 
beyond the line $\bar{\rho}_c$. Then we let 
\beq
\mcA(Y,\rho, p) = \mcA(Y,\bar{\rho}_c, p) + \bar{\rho}_c(Y,p) - \rho, \,\,\,\, 
\mathrm{for} \,\, \mathrm{all} \,\,\,  \rho \leq \bar{\rho}_c(Y,p) \; .
\label{boundary1}
\eeq
This is 
similar to the behavior observed in  the BK equation.

Another choice is to simply freeze the gluon distribution at the critical point, 
\beq
\mcA(Y,\rho, p) = \mcA(Y,\bar{\rho}_c, p), \,\,\,\, 
\mathrm{for} \,\, \mathrm{all} \,\,\,  \rho \leq \bar{\rho}_c(Y,p) \; .
\label{boundary2}
\eeq

Once the boundary has been implemented, one can extract the saturation momentum 
from the evolution.  $Q_s$ can be defined as a line of constant density. One can 
define $\rho_s$ as any line of constant density between $\rho_c$ and 
$\bar{\rho}_c$. The precise choice affects only the normalization of $Q_s$. 

We will investigate both choices of the boundary condition. This is because one does not really know the details of the non-linear dynamics in the CCFM, and we want to check the robustness of the obtained results with respect to the change of the boundary condition.

\section{Numerical results in the presence of the saturation boundary}
\label{sec:results}
\subsection{Fixed coupling results}

\begin{figure}
\begin{center}
\includegraphics[angle=0, width=0.47\textwidth]{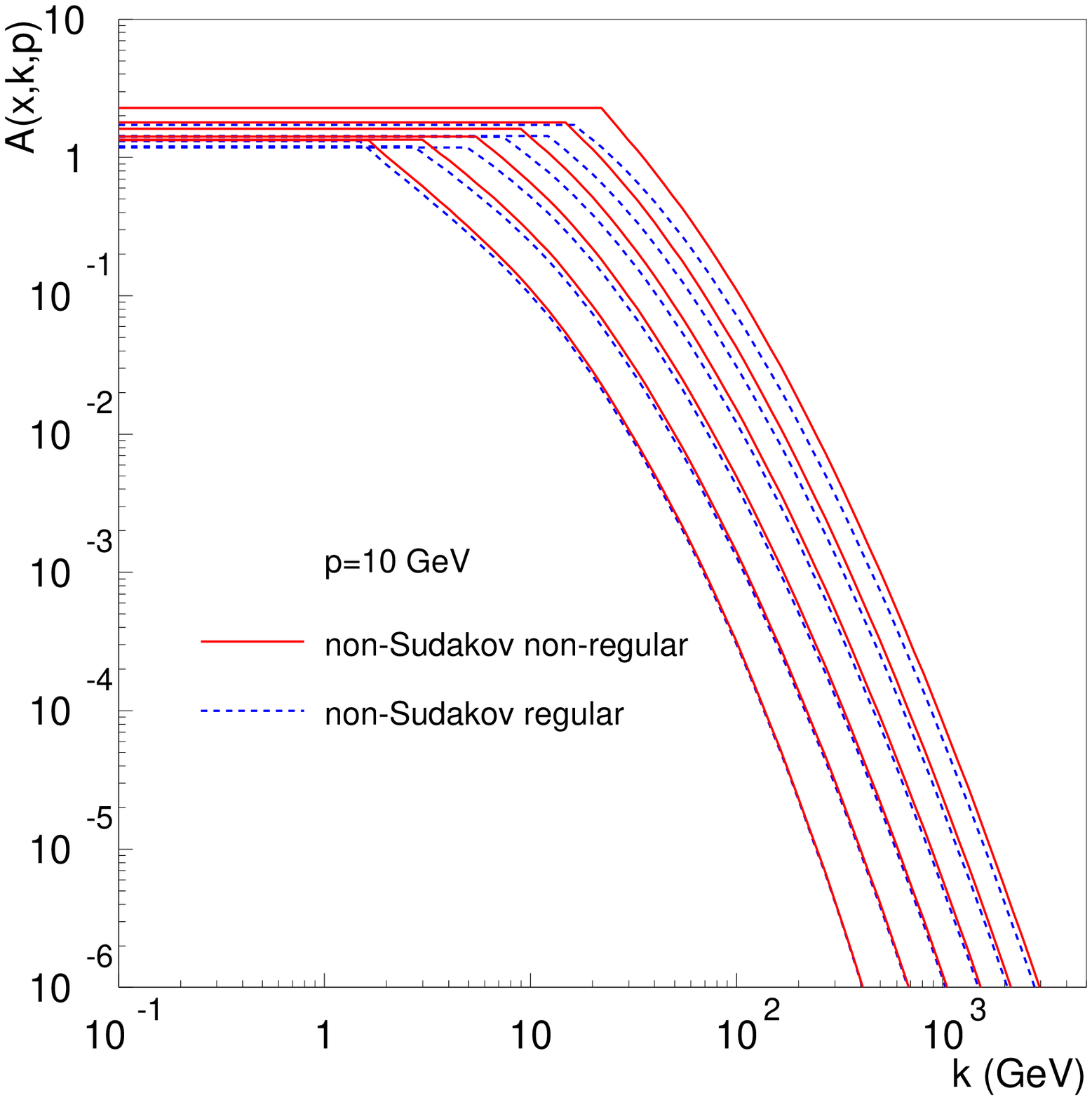}\hfill
\includegraphics[angle=0, width=0.47\textwidth]{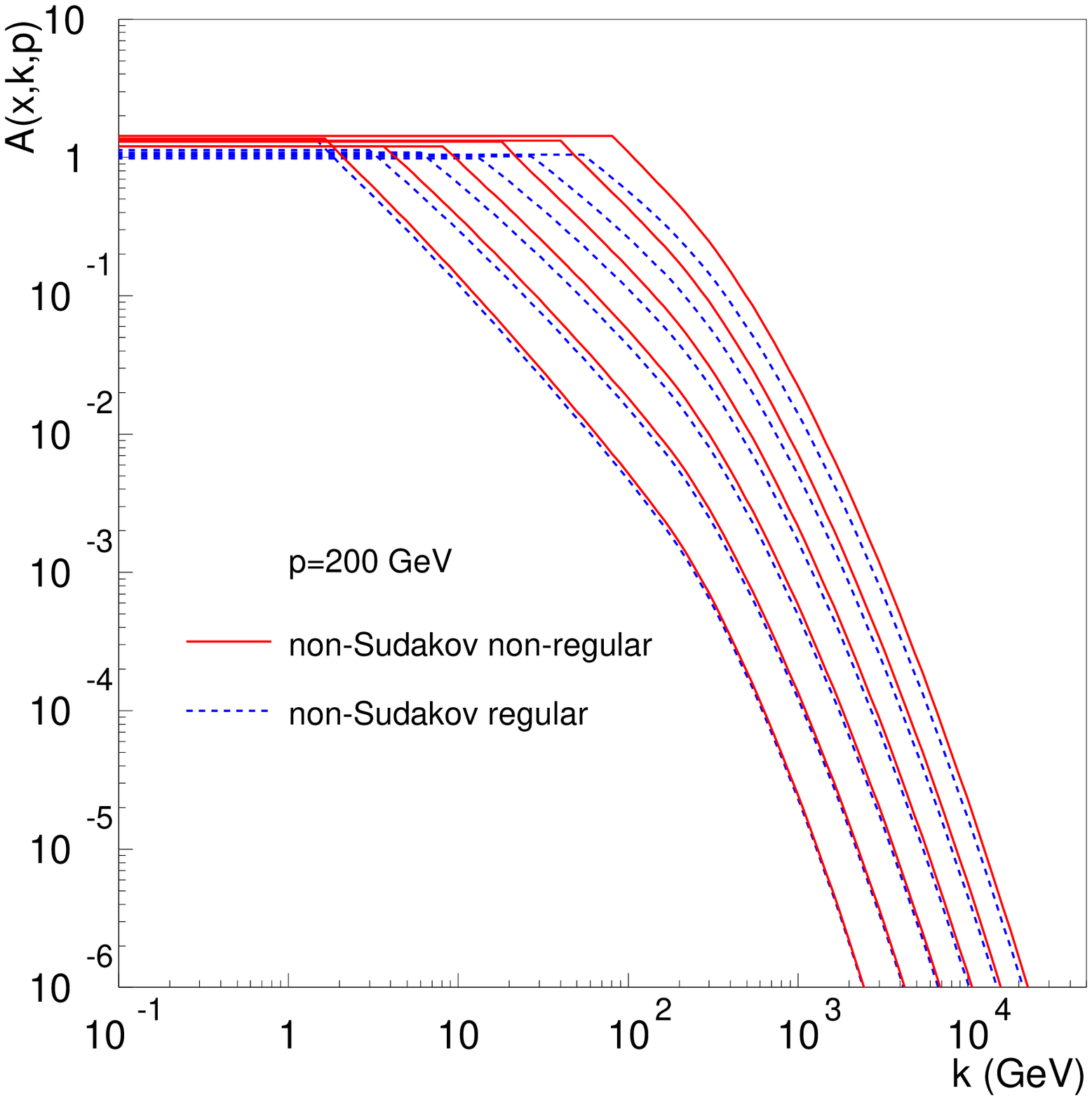}
\end{center}
\caption{Solution to the CCFM equation \protect \eqref{eq:ccfminteq2} with saturation boundary.  Curves increasing in magnitude are for rapidities $Y=4,6,8,10,12,14,16$ respectively.
Value of the external scale $p=10 \; {\rm GeV}$ (left plot) and $p=200 \; {\rm GeV}$ (right plot). Dashed blue lines correspond to the form-factor \protect \eqref{JSnonsud}; solid red curves correspond to the form factor \protect \eqref{eq:nonsud}}
\label{fig:absfxd}
\end{figure}

In figure~\ref{fig:absfxd} we show the solution to the CCFM with the saturation boundary 
implemented in the form of freezing as in \eqref{boundary2}. Here, the coupling is fixed
$\abar=0.2$. We compare solutions obtained using \eqref{eq:nonsud} and \eqref{JSnonsud} respectively. We see that once the saturation effects are included, the differences between the solutions using the different form-factors are not that large, as the sensitive infrared region is regularized by the saturation boundary.  The dependence on $p$
can be examined by comparing the left and right plot in figure~\ref{fig:absfxd}. We observe
that,  similarly to the linear case,  the distributions change slopes when $k$
become comparable to the external scale $p$. For $k > p$, the gluon distribution is effectively cut off by the scale $p$.

\begin{figure}
\begin{center}
\includegraphics[angle=0, width=0.47\textwidth]{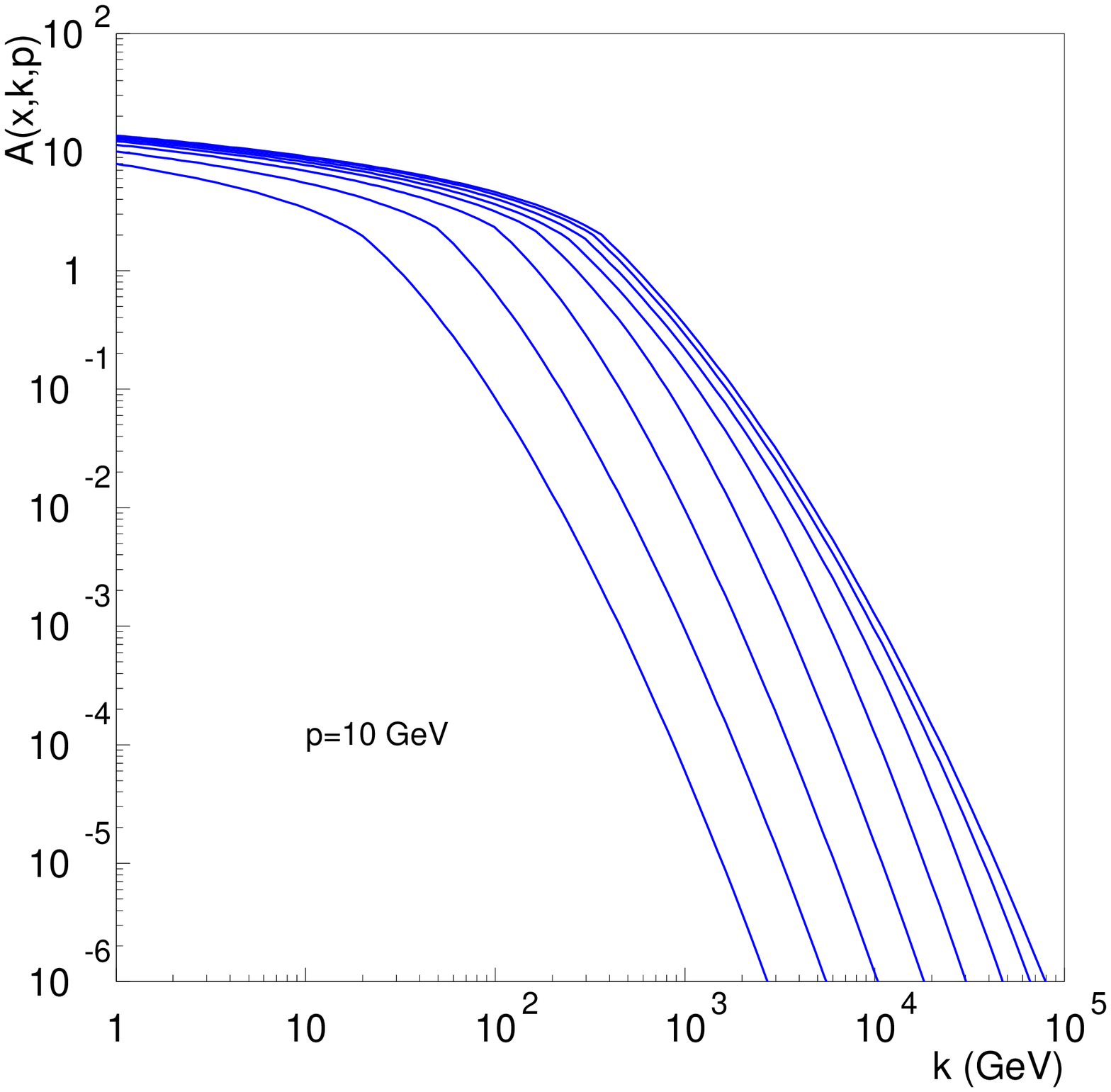}\hfill
\includegraphics[angle=0, width=0.47\textwidth]{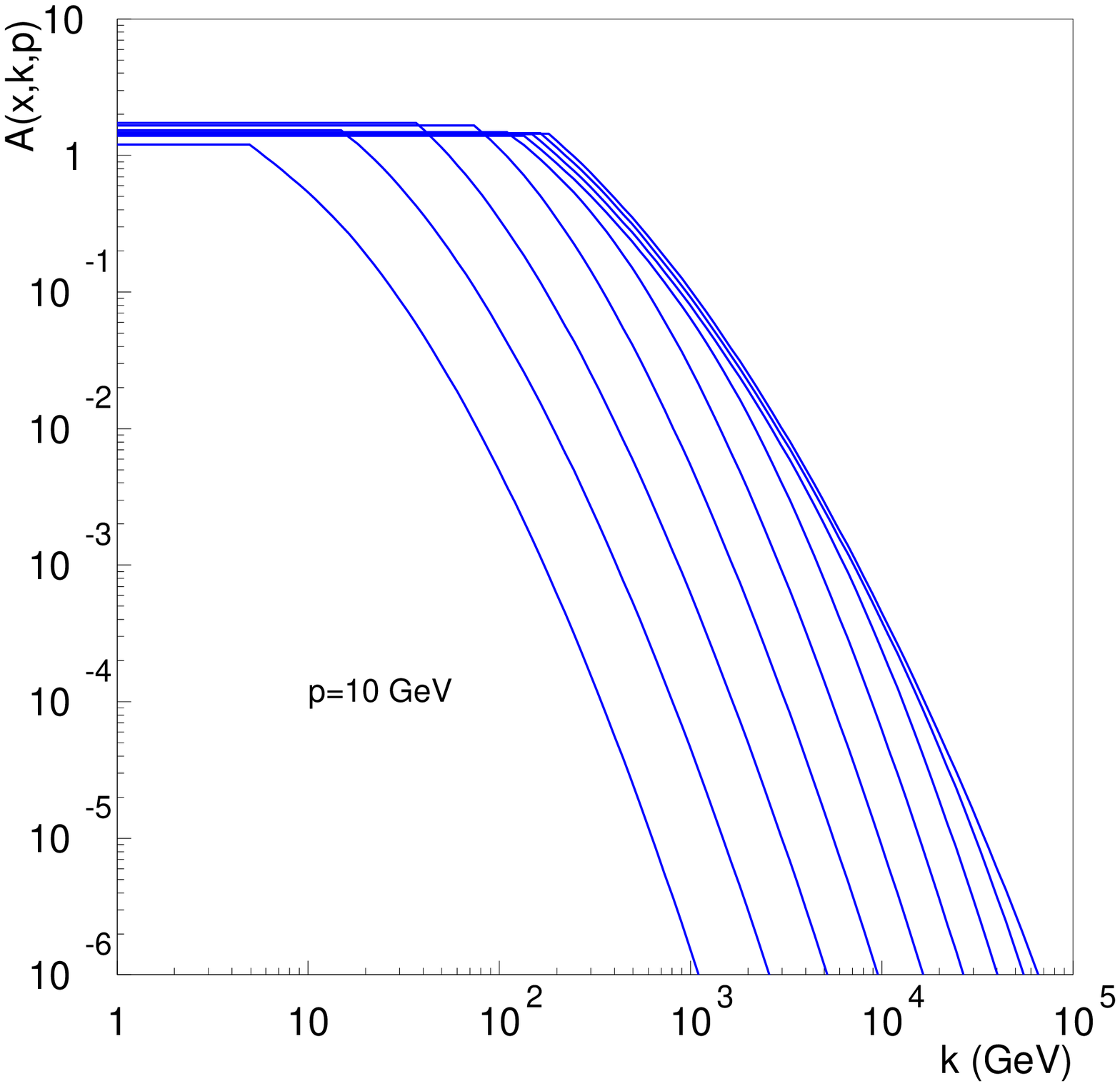}
\end{center}
\caption{Solution to the CCFM equation \protect \eqref{eq:ccfminteq2} with 
two different saturation boundaries; \protect \eqref{boundary1} (left) and 
 \protect \eqref{boundary2} (right).  Curves increasing in magnitude are for rapidities $Y=10,15,20,25,30,35,40,45,50$ respectively.
The value of the external scale is $p=10 \; {\rm GeV}$.}
\label{fig:absfxd_y50}
\end{figure}
In order to analyze the asymptotic behavior of the solutions we extend the rapidity range up to 
$Y=50$. This is shown in figure~\ref{fig:absfxd_y50}, where we use the  non-Sudakov form-factor from equation~\eqref{eq:nonsud}.  We observe the interesting effect that the front of the solution no longer travels to higher momenta with uniform 'speed' governed by the saturation scale, but that it rather stalls at high rapidities. 
The origin of this behavior is rooted in the presence of the additional scale $p$. As the rapidity grows, the solution diffuses to the higher values of the transverse momenta $k$. 
At the same time the saturated region grows also with energy. The unsaturated region
is eventually pushed to the region where $k \gg p$, which is then cut off by the angular ordering condition. Therefore the phase space for new emissions is squeezed, due to the presence of two scales $Q_s$ and $p$. 
 We have done the analysis with two different boundaries \eqref{boundary1} (logarithmic growth at low momenta)
and \eqref{boundary2}  (freezing) and it is clear that the front stops to diffuse to higher momenta in both cases.
The stalling of the front therefore is a robust feature of CCFM with saturation as it does not depend on the way the boundary is implemented.
We note that the complete stalling of the front (in the case presented in figure~\ref{fig:absfxd_y50}) occurs for momenta $k \sim 10^2 \; {\rm GeV}$, which is order of magnitude larger than the $p \sim 10$. In consequence, the saturation scale
$Q_s$ starts to behave differently with rapidity in this regime. In particular the value of $\ln Q_s$ will no longer scale linearly with rapidity but slower than that, and eventually the saturation scale will 'saturate' itself at large $Y$. The complete analysis of the behavior of the saturation scale is presented in the later section.


\subsection{Running coupling results}


\begin{figure}[t]
\begin{center}
\includegraphics[angle=0, width=0.47\textwidth]{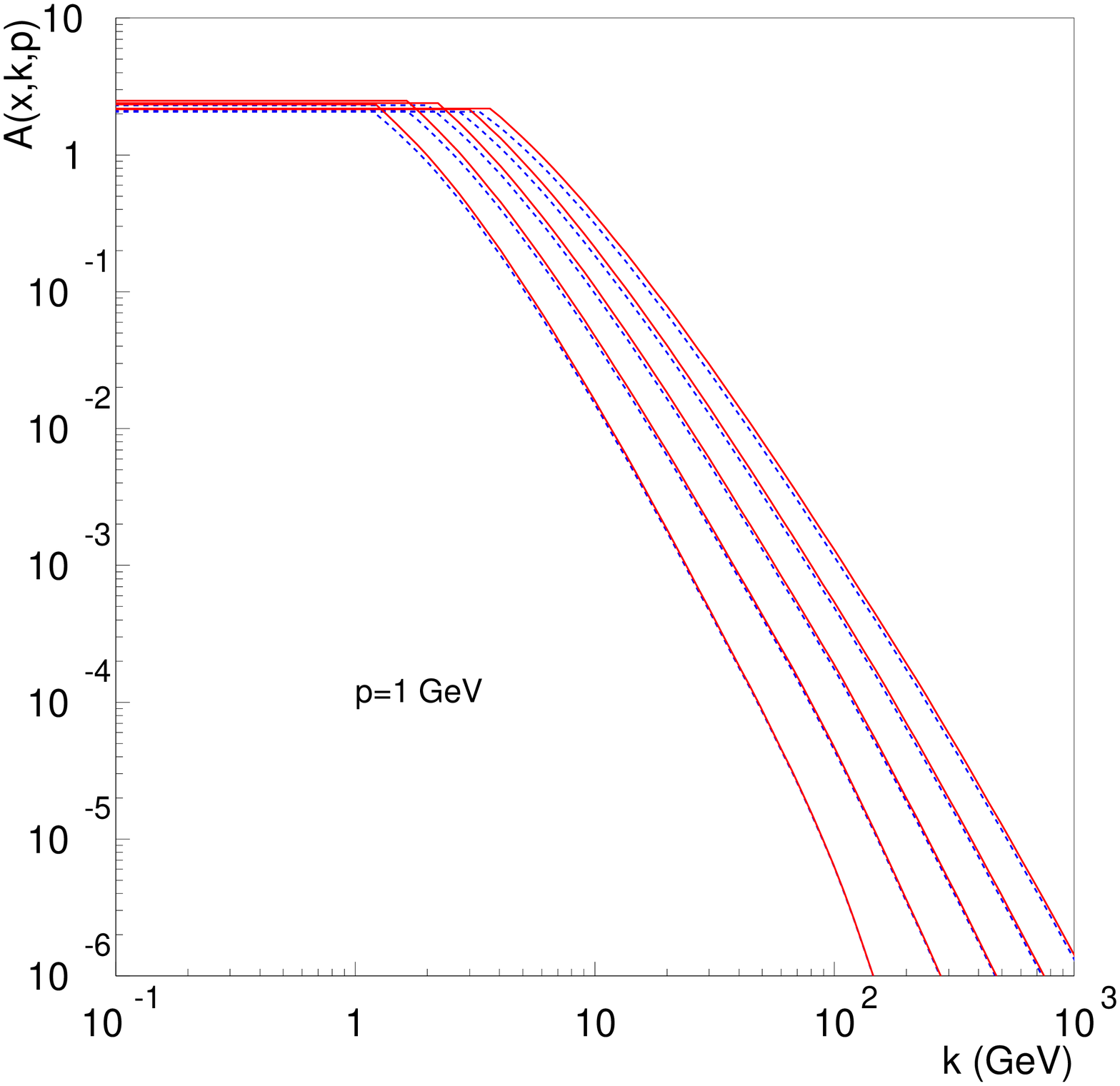}\hfill
\includegraphics[angle=0, width=0.47\textwidth]{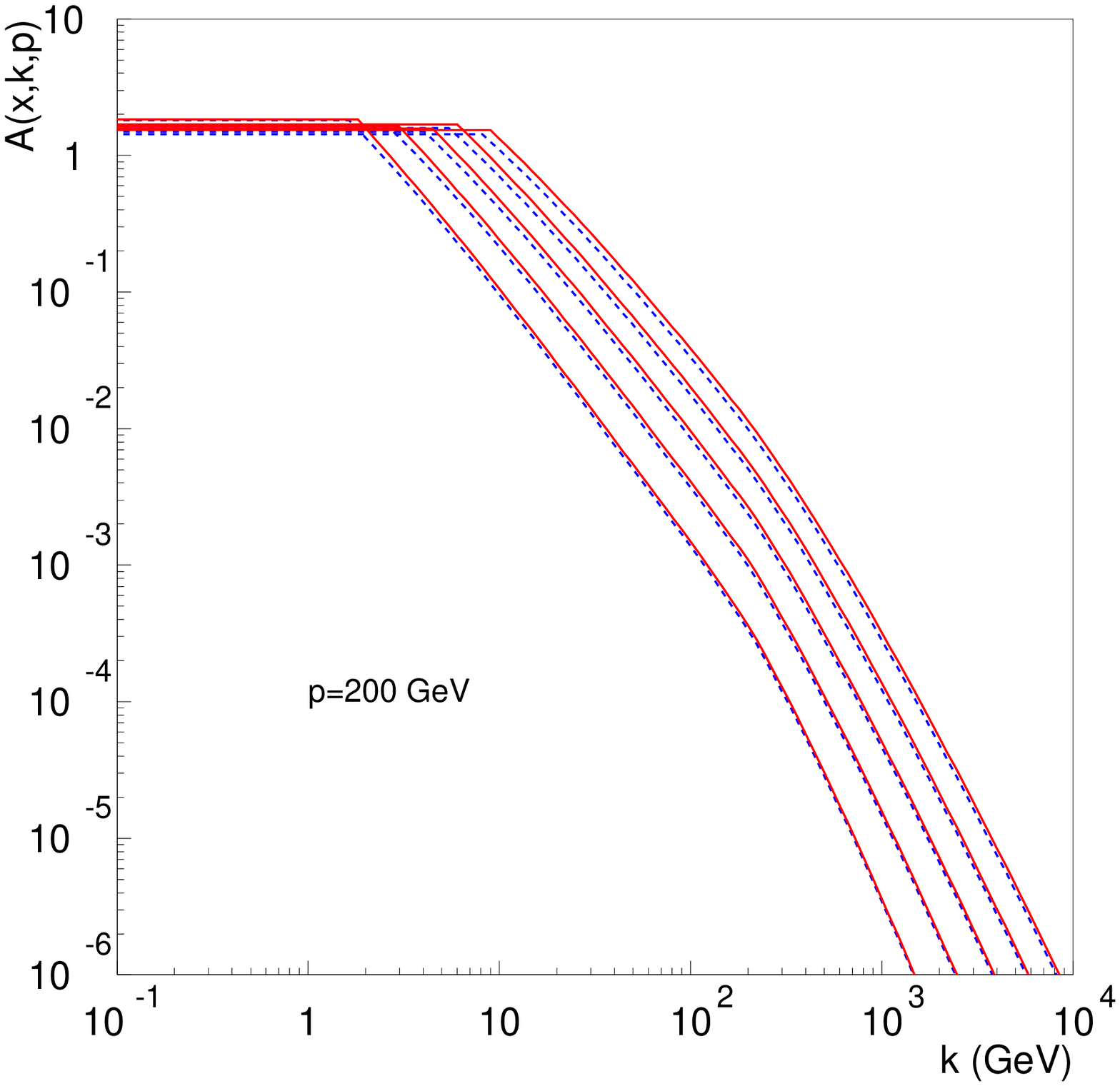}
\end{center}
\caption{\label{fig:ccfmabs1} The gluon distribution in the presence of the 
saturation boundary \protect \eqref{boundary2} for \protect $Y=6, 8, 10, 12, 14$. The solid lines are obtained 
using the form factor \protect \eqref{eq:nonsud} while the dashed lines are obtained using \protect \eqref{JSnonsud}.
Left: \protect $p=1$ GeV. Right: \protect $p=200$ GeV.
}
\end{figure}

Let us now turn the case of a running coupling. 
In figure \ref{fig:ccfmabs1} we show the different results obtained using the two different 
form factors, \eqref{eq:nonsud} and \eqref{JSnonsud}. Here, the boundary condition 
\eqref{boundary2} is used, and $k$ is used as scale in $\abar$. As in the fixed coupling case we 
see that the differences are much smaller than in the linear evolution, and almost 
identical results are obtained. In particular we have checked that the small difference between 
the results stays the same up to very large values of $Y$ (far beyond phenomenologically relevant 
values). We find that, apart from the region of smallest $Y$ in the figures, 
the difference is a normalization factor independent of $p$ and $Y$.  
Consequently the different form factors lead to almost identical saturation scales 
which we shall look more closely at in the next section. 

Since in the running coupling case the evolution is generally slowed down, 
one needs to go to higher values of $Y$ to see the stalling of $\mcA$. For larger 
$p$ in fact, the $Y$ values required for the stalling are 
so extremely large that we have not reached them in our numerical solution. 
Starting from 
smaller $p$ values, however, we clearly see how the evolution is eventually slowed
down to a point where $\mcA$ starts to stall in a region between $\bar{\rho}_c$ 
and $\rho_c$ (\emph{i.e.} around $\rho_s$), and one can also  
see how this region then grows with $Y$. Increasing $p$ we see that one needs ever higher 
$Y$ to see this behavior (as in the fixed coupling case). 


\begin{figure}[t]
\begin{center}
\includegraphics[angle=0, width=0.6\textwidth]{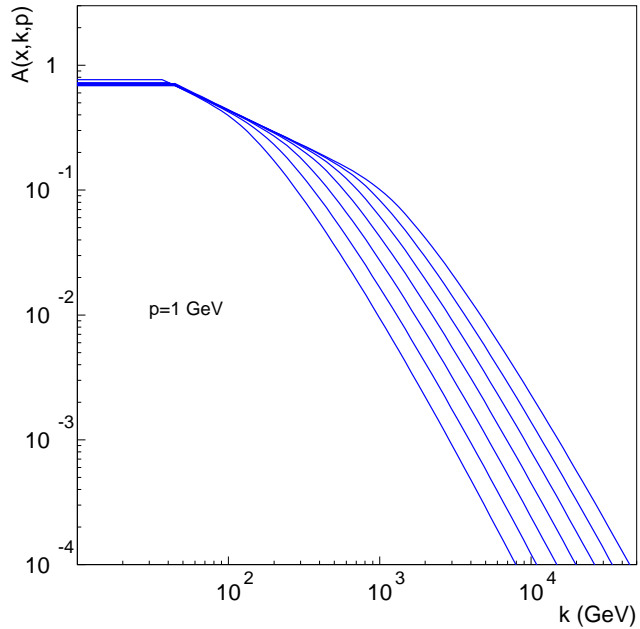}
\end{center}
\caption{\label{fig:ccfmabs2} Solution of $\mcA$ in the asymptotic region of 
$Y$ between 46 and 70 units. The non-Sudakov in \protect \eqref{JSnonsud} is used and $p = 1$ GeV. }
\end{figure}

In figure \ref{fig:ccfmabs2} we show the extension of the solution 
shown in the left plot of figure \ref{fig:ccfmabs1} (that is when $p=1$ GeV) to asymptotic values of $Y$
between 46 and 70 units. In this case only the solution using \eqref{JSnonsud} is presented, 
the solution using \eqref{eq:nonsud} looks essentially the same. Compared to the fixed coupling
case, the region where $\mcA$ stalls is much smaller, and it also grows slower. Therefore the saturation 
of the saturation scale will set in much later (to be shown in the next section). 
To be able to exactly determine at what $Y$ the stalling of $\mcA$ starts to become visible, 
we would need to know the full non-linear formulation of CCFM. The application 
of different boundary conditions seem to, however, indicate that this behavior sets in rather late, as 
manifest in figure \ref{fig:ccfmabs2}. Moreover there are additional effects not taken into 
account here (such as energy conservation)  which will further slow down the evolution.

\begin{figure}[t]
\begin{center}
\includegraphics[angle=0, width=0.47\textwidth]{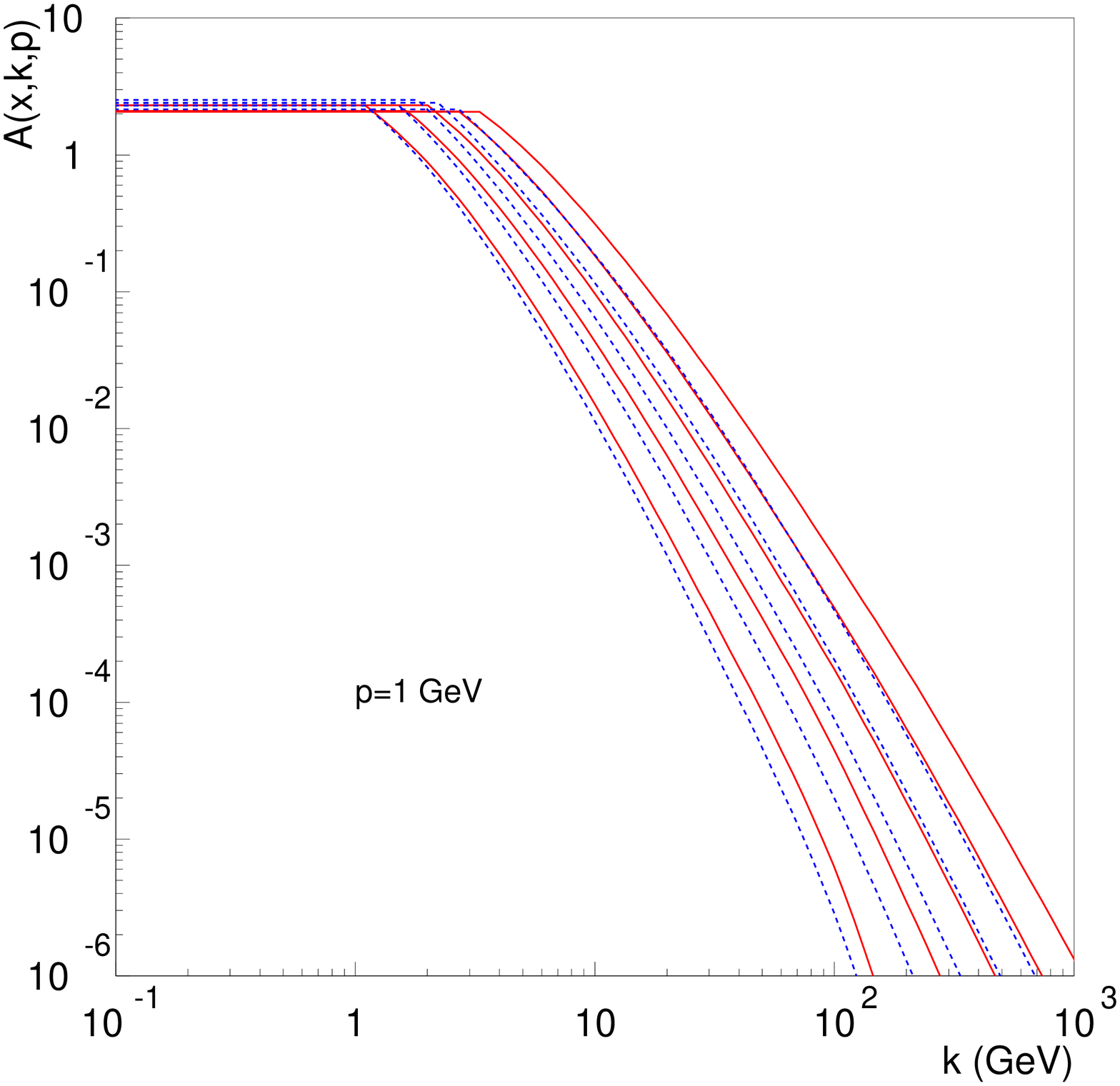}\hfill
\includegraphics[angle=0, width=0.47\textwidth]{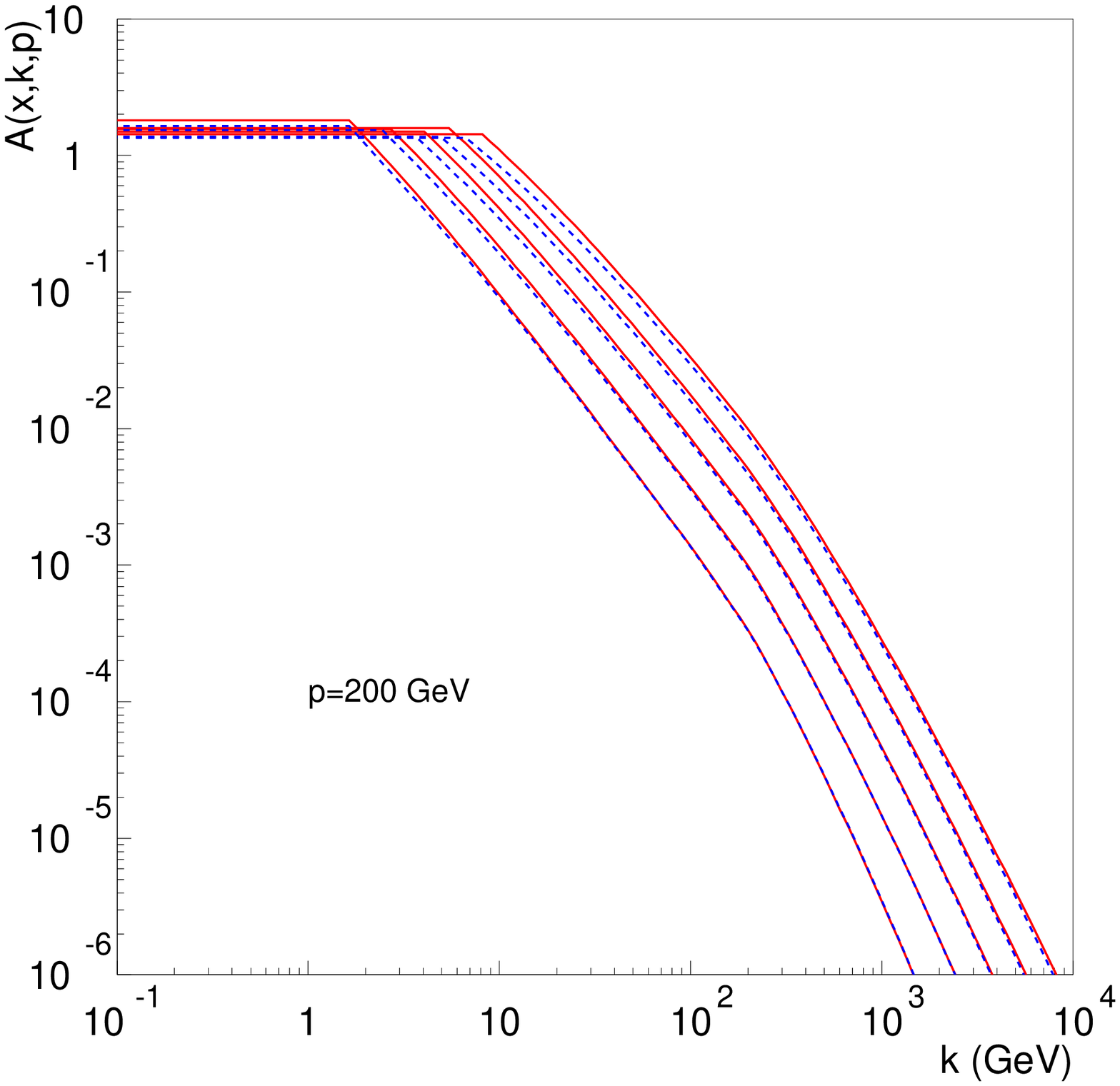}
\caption {\label{fig:ccfmabs3} Comparison of solutions using as a scale in 
$\abar$: $k$ (solid lines), $q$ (dashed lines) for $Y=6, 8, 10, 12, 14$. 
Left: $p=1$ GeV. Right: $p=200$ GeV. 
}
\end{center}
\end{figure}

Let us now turn to the implementation of different scales in $\abar$.
In the linear case we observed the fact that generally the choice of $q$ 
gives a slower evolution compared to the default choice $k$.  
In figure \ref{fig:ccfmabs3} we show the comparison between the results 
obtained from the two different scales.
For the larger $p$ values the differences between the two choices of scales are very small.
For smaller values of $p$ the differences are more pronounced, the solution with scale $q$
being consistently smaller than with the choice of scale $k$. As explained earlier in section 
\ref{sec:lindiffrun}, this effect is due to the stronger suppression from $\Delta_{ns}$ 
when $q$ is selected as scale.


This, however, brings us to another point. Namely the fact that one should actually 
not allow any contribution to the non-Sudakov form factor from the region which
lies outside the region set by the maximal angle $\bar{\xi}$, or equivalently 
by $p$. To understand this, remember the derivation of $\Delta_{ns}$ shown in figures \ref{fig:nonSud} 
and \ref{fig:nonSud2}. The regions $B_i$ in that case were bounded by the maximal angle $\bar{\xi}$.
When $k \gg p$, it can be that region $B_i$ partly, or completely, lies outside 
the region bounded by $\bar{\xi}$. In that case also parts of region $C_i$ will be located 
outside the allowed region (however, $C_i$ cannot lie completely outside 
the allowed region). In reality, however, one should not include any contribution
from those parts of $C_i$ and $B_i$ (actually since $k > q$ only figure 
\ref{fig:nonSud} and the corresponding regions $B_1$ and $C_1$ are relevant). Thus one
should in equation \eqref{eq:nonsud1st} actually only include the part of region $C_1$ which is bounded by $\bar{\xi}$.
By not doing this one is generally overestimating the suppression from $\Delta_{ns}$ 
when $k \gg p$. The non-Sudakov form factor will then be modified, and it will 
depend not only on $k$, $q$ and $z$, but also on $p$ and the previous values of $z_i$ 
in the chain. Unfortunately this also implies that it is no longer possible 
to write an integral equation as in \eqref{eq:ccfminteq1} or \eqref{eq:ccfminteq2}
because one needs to keep track of all $z_i$ values in each $\Delta_{ns}$ during the 
evolution (this is because $\Delta_{ns}$ then depends not only the vertex $k' \to k+q$ 
but also on earlier vertices). One can only write down the gluon distribution in an
unfolded form, explicitly summing it over all possible chains. 

Thus when using the scale $q$ instead of $k$ we do find some differences in the evolution. 
On the other hand, for reasonable values of $Y$ these differences are rather small, and for 
$p$ values around 10 GeV and higher they are negligible. We will also see in the next section
that the respective saturation momenta are quite similar, despite the noted differences. 
We have also noted that when $k > p$ one should actually modify $\Delta_{ns}$ to be 
consistent. However, the implementation of this modification is probably in practice only 
feasible  in a MC simulation where one can 
keep track of the history of the entire gluon chain.

\subsection{The saturation scale}

\emph{Fixed coupling results} \\

From the solutions to the gluon distribution one can extract the saturation scale $Q_s$. 
We have already commented on the definition of $Q_s$ in section \ref{sec:numerics}. 
The exact value of the constant defining $Q_s$ only affects 
its normalization which we cannot control anyhow, the $p$ and $Y$ 
dependences are independent of the value of this constant (of course 
the constant should not be varied over many orders of magnitude). 


\begin{figure}[t]
\begin{center}
\includegraphics[angle=0, width=0.47\textwidth]{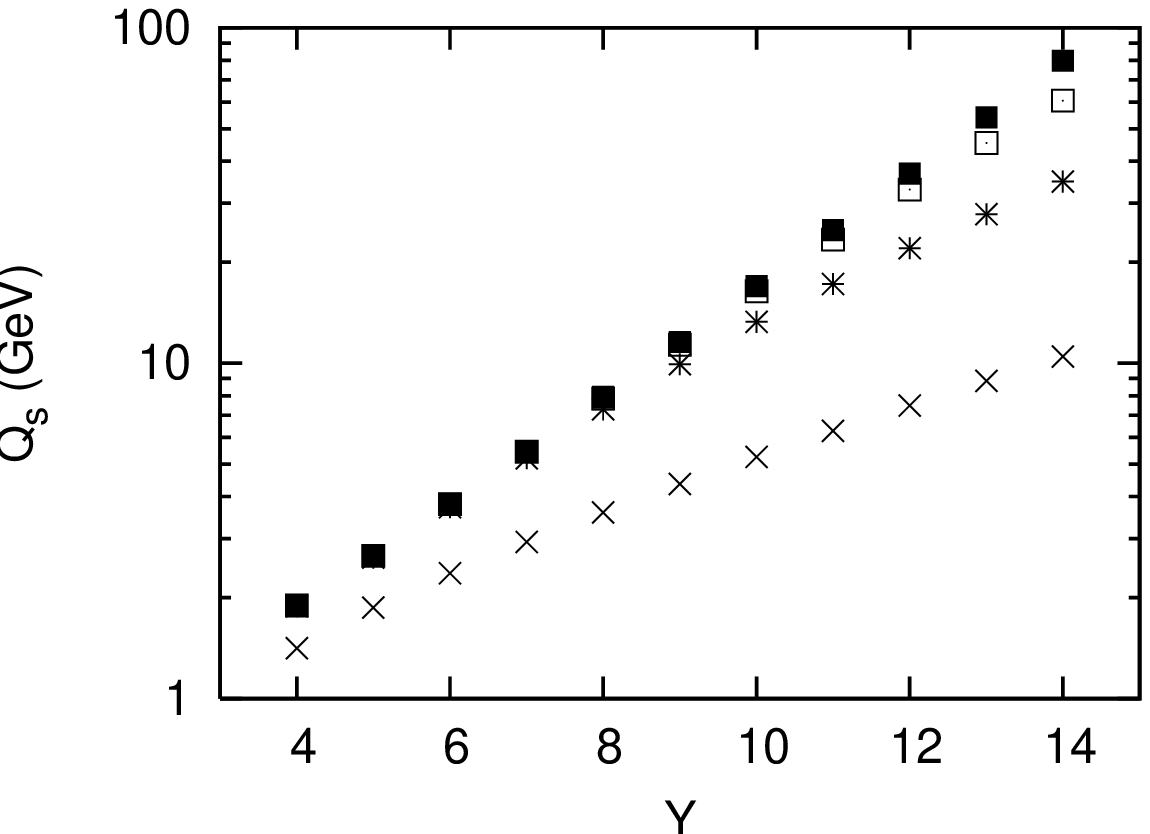}\hfill
\includegraphics[angle=0, width=0.47\textwidth]{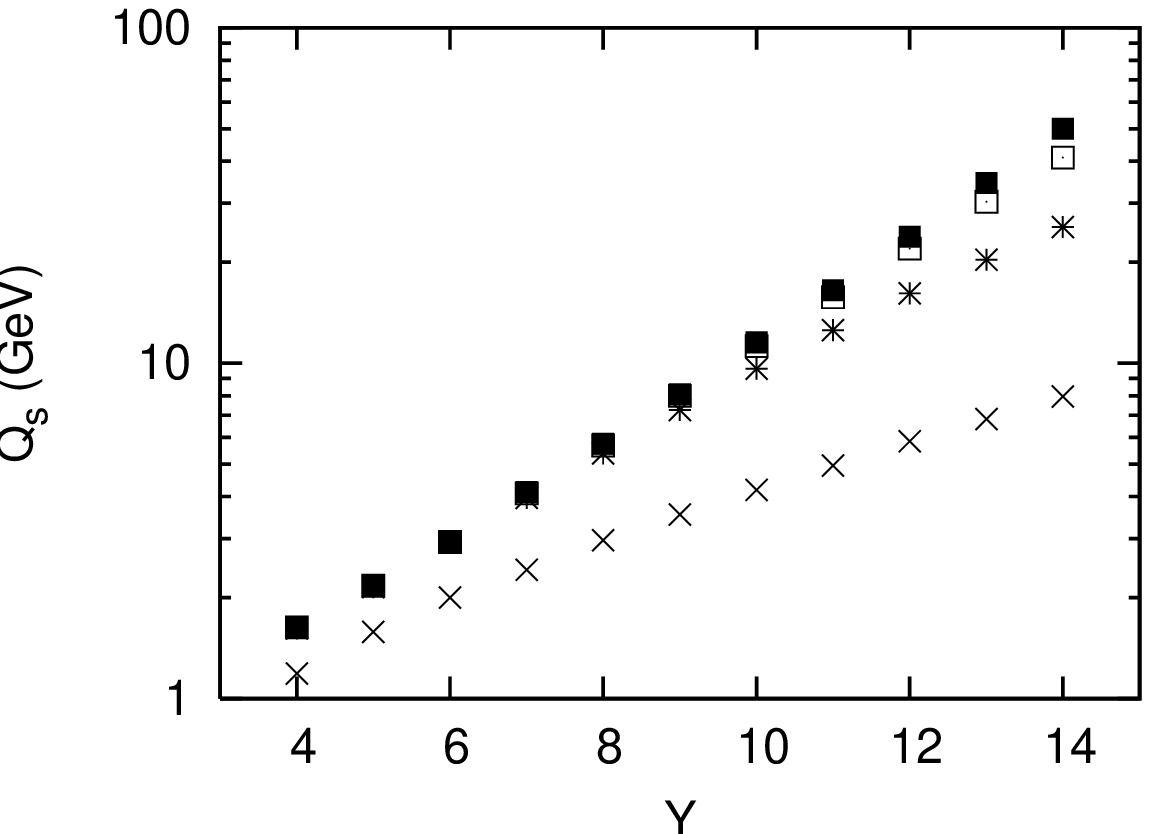}
\end{center}
\caption{\label{fig:Qsatfa} The saturation momentum $Q_s$ for $p = 1, 10, 40$ 
and 200 GeV respectively (from bottom to top) as a function of $Y$, for fixed coupling $\abar =0.2$
and using the non-Sudakov \protect \eqref{JSnonsud}. Left: Using boundary condition 
\protect \eqref{boundary1}. Right: Using boundary condition 
\protect \eqref{boundary2}
}
\end{figure}

In figure \ref{fig:Qsatfa} we show the results for the two different 
boundary conditions \eqref{boundary1} and \eqref{boundary2} 
when the coupling is fixed and the non-Sudakov in 
\eqref{JSnonsud} is used.
In both cases we find a behavior consistent with the expectation in 
\eqref{eq:satscale1} which holds as long as $Q_s \leq p$. As $Q_s$ grows 
above $p$, as is most visible in the case when $p=1$ GeV in figure \ref{fig:Qsatfa}, we see that 
the growth of $\rho_s=\ln Q_s^2/Q_0^2$ on rapidity  becomes non-linear. The $Y$ range shown in the 
figure is similar to what is  accessed at the LHC. The absolute values 
of $Q_s$ should not be considered as realistic estimates of what $Q_s$ 
would be at the LHC (a fixed coupling calculation is not so relevant phenomenologically anyway).  
The two different boundary conditions give rise
to slightly different $Y$ dependences for $Q_s$ for each given $p$, 
but the differences are rather small. 


\begin{figure}[t]
\begin{center}
\includegraphics[angle=0, width=0.6\textwidth]{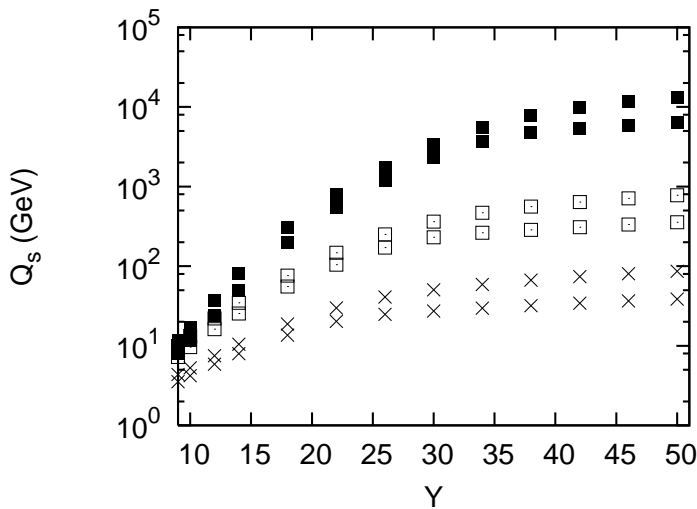}
\end{center}
\caption{\label{fig:Qsatfa2} The saturation scale from figure \protect \ref{fig:Qsatfa} 
extended to very large $Y$ in order to clearly see the saturation of $Q_s$. 
The three set of curves are for $p=1, 10$ and $200 \; {\rm  GeV}$ respectively (from bottom
to top) and the two different set of points within each set are for the 
boundary conditions \protect \eqref{boundary1} and \protect \eqref{boundary2}.
}
\end{figure}

The rather dramatic change in the $Y$ dependence of $Q_s$ is visible in figure 
\ref{fig:Qsatfa2} where we extend the results from figure \ref{fig:Qsatfa} to much larger
$Y$ (far beyond experimentally accessible values). As we see, the general shape of $Q_s$
does not depend on the precise boundary being used. We see that once $Q_s > p$, the dependence
of $\rho_s$ on $Y$ becomes non-linear, and eventually it saturates to some value depending 
on $p$. Of course we cannot say exactly whether there exist a given constant $C$ 
such that $Q_s \leq C$ as $Y \to \infty$ or 
whether $Q_s$ keeps growing, albeit extremely slowly.
Nevertheless, it is clear that, the growth of $Q_s$ on rapidity saturates and that the dependence of $\rho_s$ on 
$Y$ is highly non-linear.   
Interestingly, we observe that there exists a certain scaling in the sense that once $Q_s$ 
has saturated, the ratio between the values for two given values of $p$ is independent of $p$. In other
words, we find that the ratio $Q_s(Y,p)/p$ is a constant independent of $p$, once $Q_s \gg p$.  
The value of this constant depends on the precise boundary condition used, for example 
for the boundary in \eqref{boundary1} we find $Q_s(Y,p)/p \approx 90$ when $Y$ is very large, 
while for the boundary in \eqref{boundary2} we instead get $Q_s(Y,p)/p \approx 40$.


\begin{figure}[t]
\begin{center}
\includegraphics[angle=0, width=0.47\textwidth]{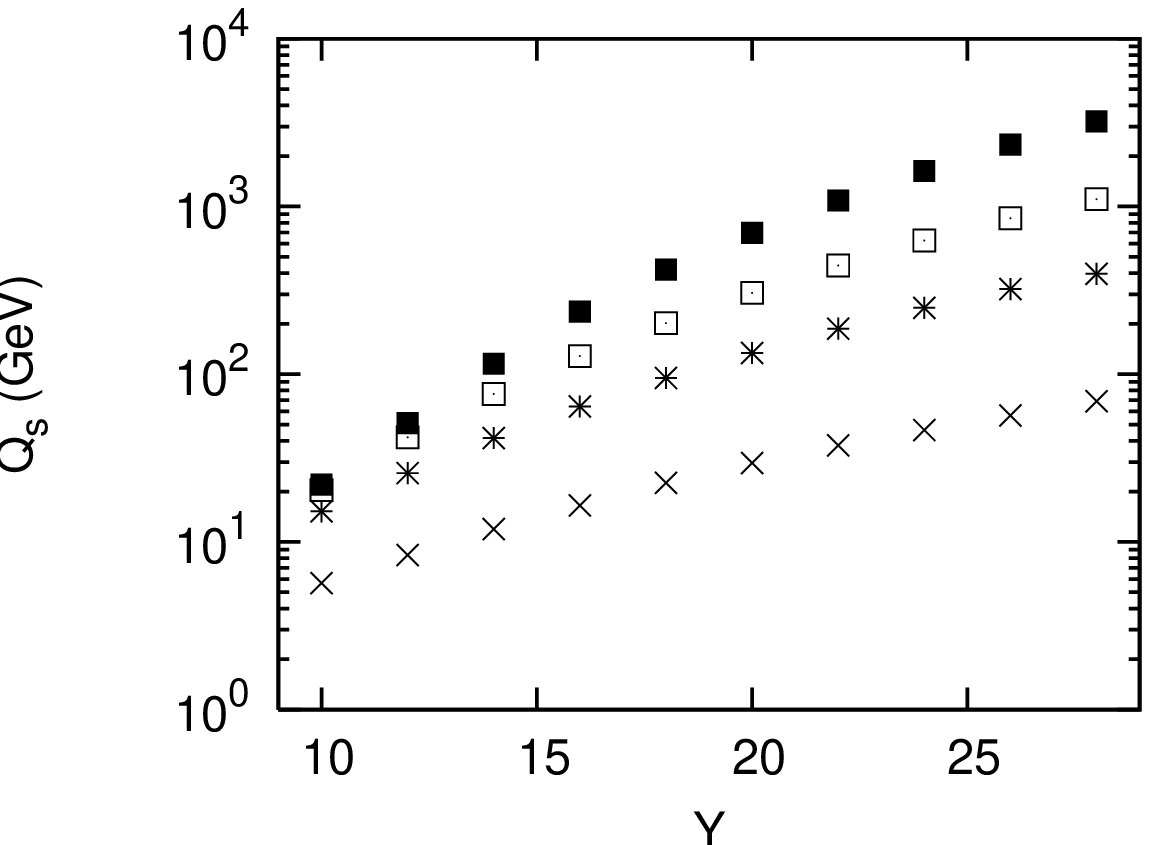}\hfill
\includegraphics[angle=0, width=0.47\textwidth]{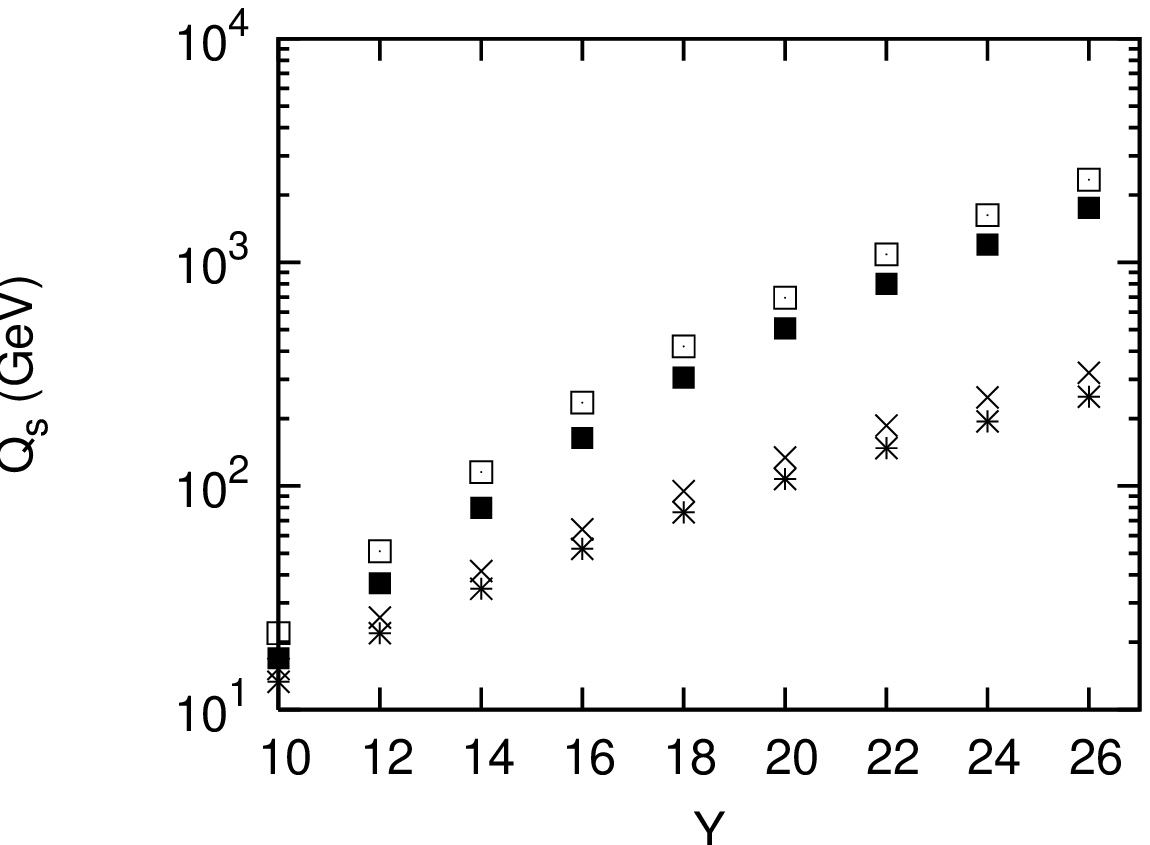}
\end{center}
\caption{\label{fig:Qsatfa3} Left: $Q_s$ obtained using the non-Sudakov in \protect
\eqref{eq:nonsud} and the boundary condition \protect \eqref{boundary1} for \protect $p=1, 10, 40$
and \protect $ 200 \; {\rm GeV}$ respectively (from bottom to top). Right: Comparison with the 
results from the same boundary condition but using \protect \eqref{JSnonsud} as 
non-Sudakov for \protect $p=1$ and $200 \; {\rm  GeV}$.  
}
\end{figure}

Next we investigate the  results obtained using the non-Sudakov in 
\eqref{eq:nonsud}. In this case since the normalization of $\mcA$ is larger, the 
normalization of $Q_s$ will also be larger. We only show the results using the 
boundary condition \eqref{boundary1}, the results for \eqref{boundary2} are similar. 
The plots for $Q_s$ are shown in figure \ref{fig:Qsatfa3} and we find exactly the same 
type of behavior as above. This is further seen from the right figure where the comparison
to the above results are shown. As can be seen, the energy dependence of $Q_s$ is the same
regardless of the non-Sudakov chosen, and we find the difference in normalization to 
be a factor roughly 1.35, regardless of the value of $p$. 
\newline

\emph{Running coupling results}
\\

It is well known that the growth of $Q_s$ is slowed down by the 
running of the coupling. Physically it is of course consequence of the fact that the
evolution leads to the diffusion towards higher transverse momenta
where the coupling is much smaller.
  In BFKL, the linear dependence 
$\rho_s \simeq \lambda Y$ at fixed coupling is for a running coupling replaced
by $\rho_s \simeq \sqrt{2\lambda_0 b_0 Y}$ where for $N_f=0$, $\sqrt{2\lambda_0 b_0} \approx 3.26$. 
In CCFM we again expect to find a BFKL-like behavior as long as 
$Q_s \ll p$. As a consequence of the much slower growth, one needs
thus in the running coupling case much higher $Y$ values in order 
to reach the point where $Q_s > p$. This implies that the stalling of 
the growth, and the consequent saturation of $Q_s$ is delayed to higher energies. 

We start by again comparing the effects of different boundary conditions. In addition to the 
two boundary conditions presented above, we also try an additional boundary 
where we simply fix $\mcA$ to be a given constant behind the saturation front. 
In particular we choose 
\beq
\mcA(Y,\rho, p) = 1,  \,\,\,\, \mathrm{for} \,\, \mathrm{all} \,\,\,  \rho \leq \bar{\rho}_c(Y,p) \; .
\label{boundary3}
\eeq
We will with this choice further demonstrate the robustness of the method. 

\begin{figure}[t]
\begin{center}
\includegraphics[angle=0,width=0.47\textwidth]{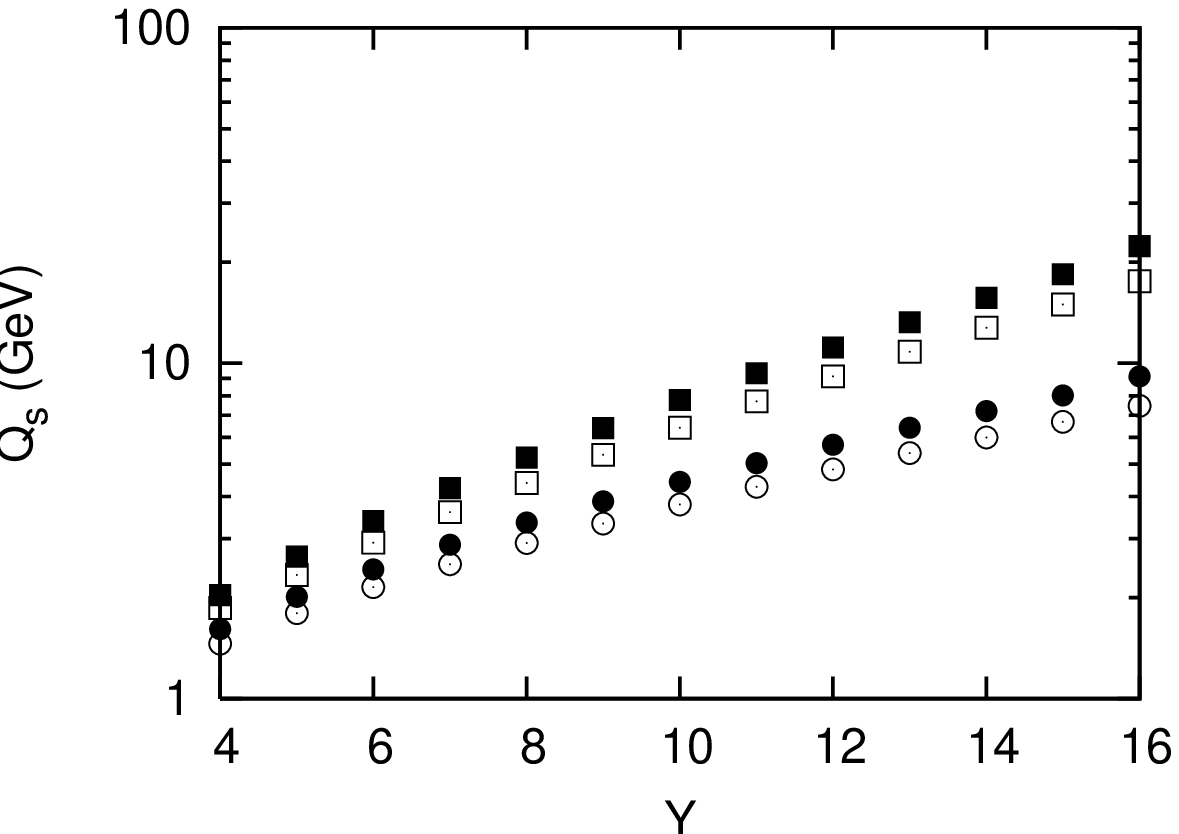}\hfill
\includegraphics[angle=0, width=0.47\textwidth]{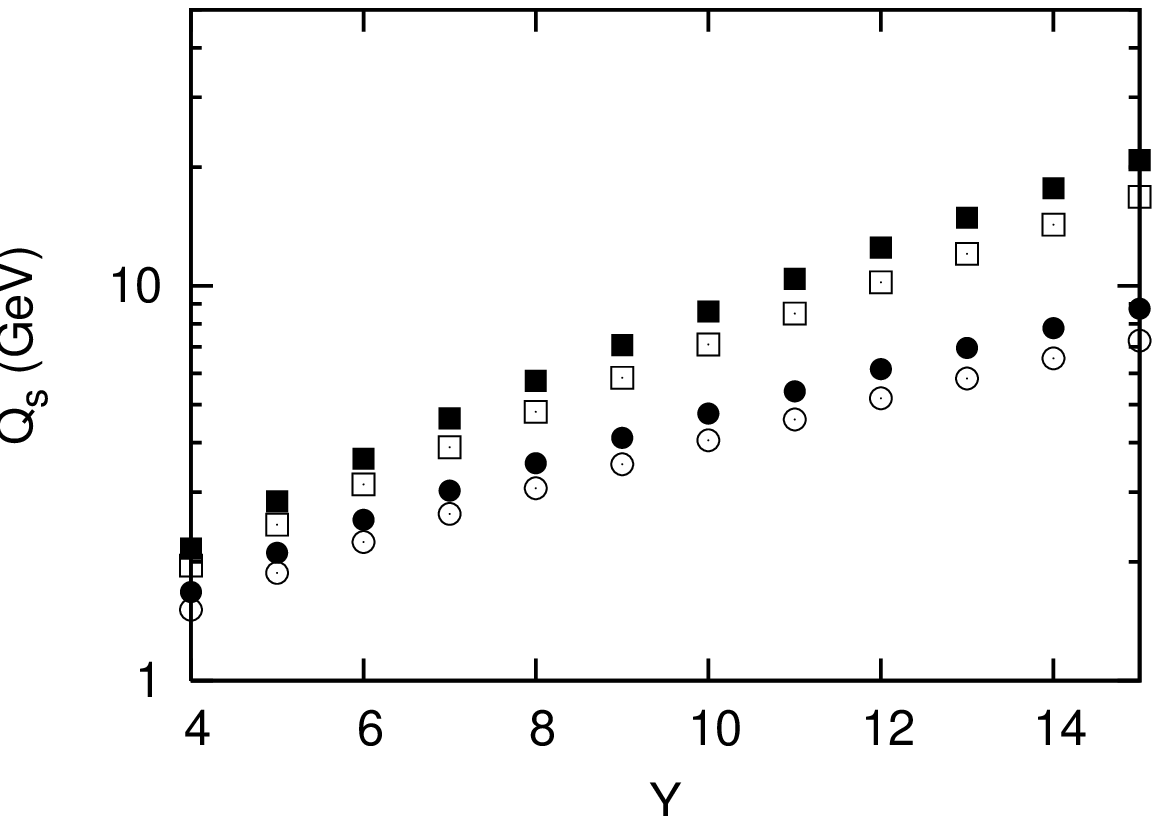}
\end{center}
\caption{\label{fig:Qsatra} $Q_s$ for different boundary conditions and $p$ values 
obtained from a running coupling; \protect \eqref{boundary2} and $p= 1$ GeV (filled circles), 
\protect \eqref{boundary3} and $p= 1$ GeV (open circles), \protect \eqref{boundary2} and $p= 200$ GeV (filled squares), 
\protect \eqref{boundary3} and $p= 200$ GeV (open squares). Left: non-Sudakov \protect \eqref{JSnonsud}. 
Right: non-Sudakov \protect \eqref{eq:nonsud}. 
}
\end{figure}

\begin{figure}[t]
\begin{center}
\includegraphics[angle=0, width=0.6\textwidth]{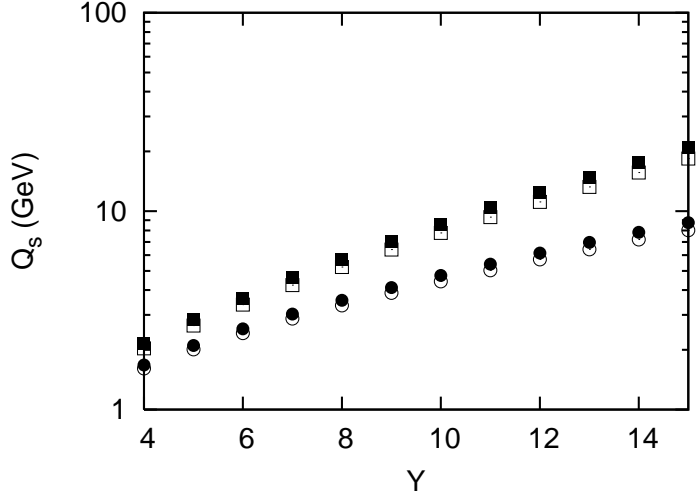}
\caption{\label{fig:Qsatra2} Comparing the results for the two different non-Sudakov form factors. 
Filled squares ($p=$ 200 GeV) and circles ($p=$ 1 GeV) are obtained from \protect \eqref{eq:nonsud} 
while the open ones from \protect \eqref{JSnonsud}. In this case the boundary condition is given by 
\protect \eqref{boundary2}. }
\end{center}
\end{figure}

In figure \ref{fig:Qsatra} we compare different boundary conditions for both of the non-Sudakov
form factors. We see that the differences are minor and that they can be removed by $p$ and $Y$ 
independent scale factors. For both cases we find a difference in scaling of around a factor 1.20. We also 
see that $Q_s$ as in BFKL grows much slower and also that, as a consequence, the dependence 
on $p$ is weaker. As in the fixed coupling case, as long as $Q_s \ll p$ we find a BFKL-like behavior, 
thus in the phenomenologically relevant $Y$ interval shown in the figure we find that 
for $p=200$ GeV, $\rho_s$ grows proportional to $\sqrt{Y}$. As $Q_s$ becomes comparable to $p$, 
however, this dependence is altered and becomes more complicated. Continuing to larger $Y$, 
one again finds that $Q_s$ eventually saturates. From figure \ref{fig:Qsatra2} we also 
see that the difference between the results for two non-Sudakov form factors are rather minor, 
and turns out to be a pure scaling, around a factor 1.10, independent of $p$
and $Y$. These results show that seemingly very different linear evolutions obtained 
from the different non-Sudakov form factors give almost identical results once saturation
is properly implemented. In this way any potential ambiguity of the formalism is 
removed. We note also that it was demonstrated in  \cite{Avsar:2009pf} that the explicit 
inclusion of the kinematical constraint  again only leads
to a different normalization, once saturation is implemented. 

\begin{figure}[t]
\begin{center}
\includegraphics[angle=0, width=0.6\textwidth]{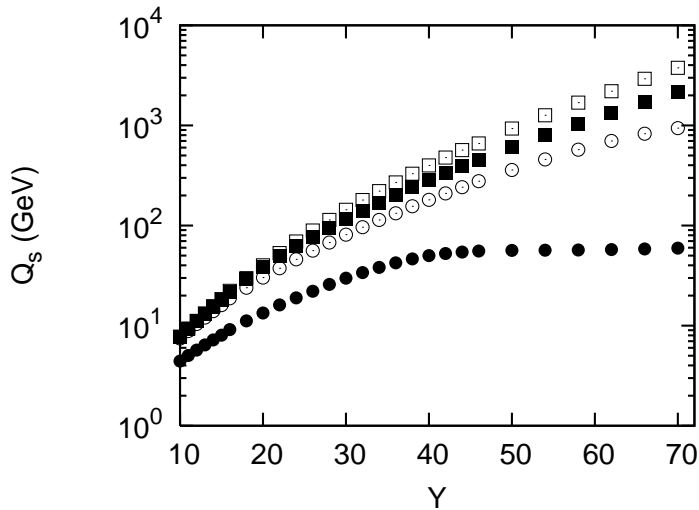}
\caption{\label{fig:Qsatra3} Asymptotic behavior of $Q_s$ for a running coupling using 
the boundary condition \protect \eqref{boundary2} and the non-Sudakov \protect \eqref{JSnonsud} for 
$p= 1, 10, 40$ and 200 GeV (from bottom to top). }
\end{center}
\end{figure}

In order to demonstrate the saturation of $Q_s$ we extend in figure \ref{fig:Qsatra3} 
the running coupling results to asymptotic values of $Y$. Once again we see the complete saturation of 
$Q_s$ for smaller $p$, but in this case one needs much larger values $Y$ to reach the stalling region. 
Furthermore we see that the scaling behavior noted above in the fixed coupling case, namely 
that $Q_s(Y,p)/p$ saturates to a number independent than $p$, is no longer true when the coupling is running. 
Additionally, as in the previous figures, we see that the dependence on $p$ is much weaker than in 
the fixed coupling case. 

\begin{figure}[t]
\begin{center}
\includegraphics[angle=0, width=0.6\textwidth]{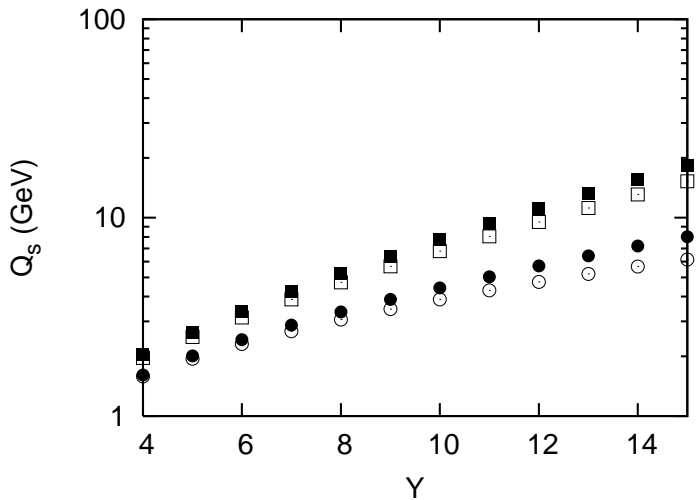}
\caption{\label{fig:Qsatra4} Left: The saturation momentum when $q$ is used as the scale 
in $\abar$. 
 The boundary condition \protect \eqref{boundary2} is used, and
$p= 1, 10, 40$ and 200 GeV (from bottom to top).  Right: Comparison with the standard 
choice of $k$ for the scale of $\abar$, for $p = 1$ and 200 GeV.
}
\end{center}
\end{figure}

Next, we extract $Q_s$ in the case when $q$ is used as scale for the running coupling. 
As we have seen from the results for the gluon distribution, the evolution for smaller $p$ is generally 
slowed down in this case, and consequently we find a slower growing saturation scale. 
The results are shown in figure \ref{fig:Qsatra4}. Since the evolution is generally slower, 
the $p$ dependence for smaller $Y$ is also weaker. For phenomenologically interesting $Y$ values, however, the differences 
are not that large. In all the results presented here for $Q_s$ one should keep in 
mind that the estimates are highly optimistic, in the sense that $Q_s$ is rather large 
already at low $Y$. From phenomenological analysis performed at HERA, we do not expect to 
find such a large normalization of $Q_s$ at the LHC. We discuss the phenomenological relevance of 
our method in the next section. 

Finally let us mention that, though not shown, we have also checked that the conclusions 
are not altered when a different set of the boundary parameters are used. 
In addition to the default choice $\Delta =2$, $c=0.4$ we have also 
used $\Delta = 1$, $c=1$ and $\Delta=5$, $c=0.1$. The different choices
lead to different normalizations of $Q_s$ but do not alter the $Y$ or $p$
dependence. The method is therefore rather general and robust. 

\section{Summary and outlook}
\label{sec:outlook}

In this paper we have analyzed the nonlinear effects in the CCFM equation using a boundary method.
We have found that, the resulting gluon distribution exhibits similar properties as in previous
BFKL/BK studies only in the region of transverse momenta lower than the scale $p$ in CCFM.
As the rapidity grows and the momenta diffuse to the higher values, the dependence on the rapidity and $p$ becomes  highly complicated, and completely novel effects are observed.
 As a result the diffusion of the transverse momenta to higher values is significantly suppressed due to angular ordering when the typical scales exceed $p$. For ultrahigh rapidities this results in the stalling of the propagation front of the solution.   

From the gluon distribution we have extracted the saturation scale in CCFM. We have found that due to the presence of the additional scale in CCFM, the saturation scale becomes also dependent on this scale. Also, the high rapidity behavior of the saturation scale is significantly modified in CCFM. For ultrahigh rapidities the saturation scale stops to grow completely, and this is related to the above-mentioned stalling of the gluon distribution.  The smaller the value of $p$, the lower are the rapidities for which deviations from the BK-type behavior of the saturation scale are observed. Analytical calculations confirm the numerical analysis on the behavior of the saturation scale.

We have investigated the cases both in  fixed and running coupling scenarios, and found that
in the running coupling case the novel features are present, yet for the higher values of rapidities.  We have also investigated the dependence of the gluon distribution on the choice of the non-Sudakov form-factors. In the case when the boundary is present there is very little sensitivity of the solutions on the different form factors, due to the fact that the nonlinear effects  basically regularize the infrared region where the differences are the largest. On the other hand, in the linear case we have found that the different forms of the form factors lead to large discrepancies of the solutions. 

Let us now comment on the possible phenomenological relevance and 
importance of the method and results presented here. 
The results obtained here indicate that the saturation of $Q_s$
occurs at extremely large values of rapidity. In the phenomenologically relevant case of the running
coupling, the rapidities needed to observe this phenomenon are about $Y=30-40$.
Such extremely large values are not accessible at present particle colliders.
One should also take into account the fact that here we only included the $1/z$ term
of the splitting function. The other, nonsingular at $z=0$ terms are important too and they are known to reduce the speed of the rapidity evolution. One also needs to properly take into account the effects of the energy momentum conservation which will further reduce the growth. 
Taking these effects into consideration we can conclude that it will be very difficult to observe
the novel effect of saturation of $Q_s$ in the near future.
One the other hand one needs to note that the exact values of the rapidities at which stalling 
occurs could depend on the details of the initial condition.  What could be however relevant phenomenologically is the  dependence of the saturation scale on the additional scale $p$
in the regime where $Q_s$ is of order $p$.  For example in DIS, in the usual saturation-based approaches one treats $Q$ and $Q_s(x)$ as completely independent scales.
The results of our analysis indicate that  saturation scale should be correlated with $Q$
when both scales are of the same order.

The interesting question is, however, whether or not it is urgent 
to introduce saturation effects for studies at LHC involving 
unintegrated parton densities. This question is relevant and important, 
as saturation affects the shape of the unintegrated distributions, 
even at higher $k$. Usually the distributions are extracted
using linear evolution equations and they 
will be used in almost every physics analysis at the LHC.
A recent study \cite{Caola:2009iy} has reported deviations from 
small-$x$ resummed DGLAP evolution in the HERA data 
that might be related to saturation effects.
Therefore we think that there is a need to analyze the possible saturation effects
in the evolution and their impact onto the LHC observables with substantial precision.

To study in more detail the relevance of saturation, one would 
first have to include additional corrections to the 
hard emissions. Obviously the soft emissions associated with the $1/(1-z)$ 
pole are important to include as well as the other non-singular terms in $z$ of the splitting function. We intend to return 
to this issue in a future publication. The soft emissions are 
included in the CASCADE event generator, and as already outlined in \cite{Avsar:2009pv, Avsar:2009pf}, 
the boundary method can be implemented in an event generator much in the same way. 
A first attempt of applying a boundary method in CASCADE was reported in 
\cite{Kutak:2008ed}. In this case, however, $Q_s$ was assumed a priori to be 
of the standard form $Q_s \sim x^{-\lambda}$ \cite{Stasto:2000er}, and it was 
then used as a cut-off in $k$. As we have seen here, $Q_s$ in CCFM depends on rapidity and $p$ in much more complicated way.
 Moreover, the power of the boundary method is that
it gives a $Q_s$ consistent with the evolution.  

A problem we see with the application of the boundary method in CASCADE is that the 
results coming from it show almost no growth at all of $\mcA$ from the 
small-$x$ evolution. The relatively steep growth in $x$ needed to fit data at HERA 
is rather provided by the $x$ dependence of the off-shell hard matrix element
and the initial condition. Obviously, if there is very little 
growth coming from the perturbative gluon cascade, then saturation 
will not be relevant. 

A similar effect is observed in  the resummed BFKL/DGLAP approaches \cite{Ciafaloni:2003ek,Ciafaloni:2003rd,Ciafaloni:2003kd,Ciafaloni:2007gf,Altarelli:1999vw,Altarelli:2001ji,Altarelli:2003hk,Altarelli:2005ni,Altarelli:2008aj}. It  was consistently found that the resummed gluon-gluon splitting function does indeed have the singularity at small $x$, but the subleading corrections cause the resummed splitting function to be very close numerically 
to NLO DGLAP in the   region $x \ge 10^{-4}$. Therefore in the kinematic regime covered by HERA the evolution of the gluon distribution is controlled by the DGLAP equations.
On the other hand it was found in earlier analysis \cite{Kwiecinski:1997ee} that when evaluating $F_2$ structure function from the $k_T$ factorization approach together with unintegrated gluon distribution, large
enhancement in the low $x$ region is  provided by the off-shell matrix element, similarly
to what was observed in CASCADE. 

A related problem was also reported in \cite{Bottazzi:1998rs} 
where some phenomenological applications of CCFM were presented. In that 
case, again only the hard emissions were included, but to simulate possible
non-leading effects, the non-Sudakov \eqref{JSnonsud} was modified
by simply changing the integration intervals, resulting in a larger 
suppression coming from the form factor (the exact motivation for this change
was given by checking that the resulting high energy saddle point more or less
agree with the one extracted from NLL BFKL studies). The result was that 
the small-$x$ growth turned out be much delayed again, and as discussed
in that paper this delay might be behind the fact that CCFM predictions
on forward jets seem to underestimate the growth of the jet cross section
with decreasing $x$ (for recent studies of forward jets within CASCADE
see \cite{Deak:2009ae, Deak:2009xt}). 

On the other hand, a reason why 
one has to introduce a strong suppression from additional effects, 
which then delay the small-$x$ growth, is because it is otherwise  
not possible to fit $F_2$ from the linear evolution. This is not so 
surprising since the linear evolution at small momenta gives
a very strong growth which we know must be tamed by non-linear effects. 
Even if one only performs a fit to $F_2$ for moderate values of $Q$, 
because one is integrating $\mcA$ to smaller $k$, saturation effects
are not completely negligible. One can therefore imagine that a procedure 
like that in \cite{Bottazzi:1998rs}, where the non-leading effects were
introduced by hand, easily overestimates the suppression coming 
from them when saturation effects are not properly included. On the other 
hand the non-leading pieces are more properly included in CASCADE which 
has the same problems as in \cite{Bottazzi:1998rs}. An event generator
is, however, rather complex, including many different components, and 
it is not always very transparent how big the role of the first principle physics 
in it is. Therefore we believe it will be very valuable to include 
the soft emissions in a simple numerical simulation as done here, 
so that one can more closely study the combined effects of the different
physics components in a simpler environment. This may shed light on the 
possible relevance of saturation for hard physics at the LHC (obviously for the 
possible description of the Underlying Event (UE) saturation effects such as 
multiple scatterings etc are very important. For a possible description of the 
very complex UE at the LHC one will need an entirely different approach 
to the problem, for some recent articles on this issue see \cite{Gustafson:2009qz, Flensburg:2010kq}).

\section*{Acknowledgments}
This work is partially supported by U.S. D.O.E. grant number DE-FG02-90-ER-40577, U.S. D.O.E. OJI  grant number DE-SC0002145 and   MNiSW  grant number
N202 249235.
A.M.S. is also supported by the Sloan Foundation.

\bibliographystyle{JHEP}
\bibliography{refs}

\end{document}